\documentclass[a4paper]{article}
\usepackage{jheppub}

\usepackage{mathabx}
\usepackage{mathtools}
\usepackage{appendix}
\usepackage{slashed}
\usepackage{multirow}
\usepackage{array}
\usepackage{booktabs}
\usepackage{cellspace}
\usepackage{makecell}
\usepackage{braket}
\usepackage{nicematrix}
\usepackage{comment}
\usepackage{hyperref}
\usepackage{subcaption}

\usepackage[T1]{fontenc} 

\newcommand{\pderiv}[2]{\frac{\partial #1}{\partial #2}}
\title{\boldmath Generalised hydrogen interactions with \texttt{CINCO}: a window to new physics
}

\author[1]{Martin Bauer,}
\author[2]{Javier Perez-Soler}
\author[2]{and Jack D. Shergold}

\affiliation[1]{Institute for Particle Physics Phenomenology, Department of Physics,\\Durham University, Durham, UK}
\affiliation[2]{Instituto de Física Corpuscular (CSIC-Universitat de València),\\
Parc Científic UV, c/ Catedrático José Beltrán 2, E-46980, Paterna, Spain}

\emailAdd{martin.m.bauer@durham.ac.uk}
\emailAdd{javier.perez.soler@ific.uv.es}
\emailAdd{jack.d.shergold@ific.uv.es}

\abstract{We present semi-analytic solutions for atomic transition rates in hydrogenic atoms induced by scalar, pseudoscalar, vector, axial-vector, and tensor interactions. Our results agree with quantum electrodynamics predictions to $\sim 0.005\,\%$ precision, and further allow us to calculate absorption and emission rates for axions, hidden photons, light scalars or other dark matter candidates for hydrogen and hydrogenic ions. These results can be used to inform searches for light new physics as well as in calculations relevant to searches for fifth forces or varying fundamental constants, with applications from astrophysics to laboratory spectroscopy experiments. We also provide a dedicated tool for the construction of hydrogenic transition amplitudes: ``Computation of hydrogen radial INtegrals and
COefficients'' (\texttt{CINCO}).}

\begin{document}
\maketitle
\flushbottom
\newpage

\section{Introduction}\label{sec:introduction}
Hydrogen is both the best understood quantum mechanical system and the most abundant element in the universe, making hydrogen transitions one of the most sensitive and versatile precision probes for new physics~\cite{RevModPhys.28.53,ANDERS19822363,Lodders:2003vvq}. These tests range from the analyses of the Lyman-$\alpha$ forest in astrophysics and cosmology to precision spectroscopy measurements in the laboratory. Any deviation from the predicted properties of hydrogen, which have been computed to extreme precision using quantum electrodynamics (QED), could be a smoking gun for new physics. The analysis of hydrogen atomic transition lines can provide evidence for dark matter interacting with electrons. They are also sensitive to variations of fundamental constants, such as the electron mass or the fine-structure constant. Similarly, hydrogen spectroscopy can detect a fifth force inducing changes in the Coulomb potential. 

The strength of different absorption or emission transition rates in hydrogen or hydrogenic atoms depends on the Lorentz structure of the underlying interaction. While QED is a pure vector interaction, interactions induced by new physics can in general be of scalar, pseudoscalar, vector, axial-vector or tensor type. A calculation of these \emph{generalised hydrogen interactions} has so far been missing in the literature, to the best of our knowledge.

In this paper we will present the general theory of hydrogen transitions via the absorption or emission of a particle or set of particles with zero total electric charge, extending and completing the results from several previous works that computed the rates for specific transitions (\textit{e.g.}~\cite{PhysRev.151.1189,Huang:2019rmc}).

In general, the computation of hydrogenic transition amplitudes involves three steps: first, calculating the angular integrals that enforce selection rules; second, evaluating the radial integrals, which are numerically challenging and involve products of hypergeometric functions that rapidly diverge at large radii; and third, analysing the operator structure to which the hydrogen electrons are coupled. These challenges are further amplified when properly treating the relativistic hydrogen atom, which is required at large values of the atomic number $Z$, or when the transition involves multiple Lorentz structures, in which case interference terms can arise. To that end, we have developed ``Computation of hydrogen radial INtegrals and
COefficients'' (\texttt{CINCO}), a powerful tool for quickly and precisely computing generalised hydrogenic transition amplitudes.

Our calculation also allows us to identify the origin of the power-law scaling for transitions in hydrogen-like ions with high charge for non-renormalisable interactions, and we emphasise how this can be exploited using transition spectroscopy in highly-charged ions as a powerful precision test for new physics. We demonstrate our results using the $2P_{3/2} \to 1S_{1/2}$ photon emission line, and a dark spin 1 boson that has both vector and axial-vector interactions, leading to absorption via the $1S_{1/2} \to 2S_{1/2}$ transition. 

The remainder of the paper will be structured as follows. In what remains of Section~\ref{sec:introduction}, we will review the role of atomic transitions in new physics searches, followed by a brief overview of the relativistic hydrogen atom in Section~\ref{sec:rha}. In Section~\ref{sec:rates}, we will then derive the general transition rate for a hydrogenic atom alongside an introduction to \texttt{CINCO}, before giving some examples of transition rates computed using \texttt{CINCO} in Section~\ref{sec:examples} and concluding in Section~\ref{sec:conclusions}. 

\newpage
%
\subsection{New physics in atomic transitions}
Spectroscopy with hydrogen is one of the most powerful experimental techniques to search for new forces, time-dependent variations of fundamental constants and in searches for dark matter due to scattering and absorption.

Searching for axions with nuclear excitations was first proposed by~\cite{Donnelly:1978ty}, and using atomic transitions in~\cite{Sikivie:2014lha}. As expected the axion mass range targeted in resonant nuclear transitions is $\mathcal{O}(\mathrm{MeV})$, whereas atomic transitions are sensitive to $\mathcal{O}(\mathrm{eV})$ axion masses. The proposal of Sikivie~\cite{Sikivie:2014lha} involves atoms cooled into the ground-state, such that only axion absorption populates a first excited state. These states are detected by a tuneable laser populating a second excited state that can only be accessed if the first excited state is populated, and subsequently detecting the de-excitation photons. Utilising an external  magnetic field allows to scan over different axion masses. An experimental proposal using this technique in molecular oxygen gas has unique sensitivity in the $\mathcal{O}(\mathrm{meV})$ mass range~\cite{Santamaria:2015gro}. Similar mass ranges can be probed utilising the hyperfine splitting of hydrogen~\cite{Yang:2019xdz}. A different proposal to observe directly atomic transitions from axion dark matter absorption has been published recently~\cite{Vergados:2024vky}. Axions interacting with photons can also trigger atomic transitions via axion-photon scattering as discussed in~\cite{Flambaum:2018wbu,TranTan:2018bch}. If the dark matter contains spin one vector bosons that mix with the SM photon, transitions in hydrogen atoms (or hydrogen-like ions) can be used to resonantly detect atomic transitions~\cite{Yang:2016zaz, Alvarez-Luna:2018jsb}. A discussion of resonant dark matter absorption in molecules taking into account cooperative effects can be found in~\cite{Arvanitaki:2017nhi}. Further, atomic transitions in hydrogen are one of the most important tools in astrophysics, e.g. the Lyman-$\alpha$ forest or the 21 cm hyperfine splitting line, for which axions and hidden photons have been considered as a possible explanation for the absorption dip observed by the EDGES experiment~\cite{Auriol:2018ovo}. 

Ultralight dark matter produced through the misalignment mechanism can also induce time-dependent oscillations of fundamental constants that can be tested by spectroscopy experiments looking for time-varying frequencies in atomic transitions~\cite{Fortier:2007jf, Rosenband:2008qgq, Derevianko:2013oaa, Arvanitaki:2014faa,Godun:2014naa, 2014PhRvL.113u0802H}. By extracting the sensitivity of these measurements to variations of electron, proton and neutron masses as well as the fine-structure constant, different couplings of this type of dark matter could be probed. 

New, so-called \textit{fifth forces} that act on electrons and nucleons lead to corrections to the Coulomb potential and therefore the line-spectrum of hydrogen~\cite{Salumbides:2013dua}. Measurements of isotope-shifts in which the frequency of atomic transitions of different isotopes are compared can isolate the interaction strength of the new force with protons and neutrons\footnote{Within this work, we will focus on exclusively proton-philic interactions, however.}, respectively~\cite{King:63, Delaunay:2017dku, Flambaum:2017onb, Berengut:2017zuo, Solaro:2020dxz, Door:2024qqz}. The sensitivity of spectroscopy experiments with Rydberg atoms has been reported in~\cite{Jones:2019qny}. Similarly, atomic and nuclear clocks have been considered~\cite{Brzeminski:2022sde}. Beyond searches for Yukawa-like new forces, hydrogen spectroscopy has been considered as a probe for Higgs-like forces~\cite{Delaunay:2016brc}, forces from axion-pair exchange~\cite{Bauer:2022rwf} or from neutrino-pair exchange~\cite{Dzuba:2017cas}. Similarly, atomic transitions in hydrogen are one of the most sensitive experimental probes for energy shifts induced by the change in the gravitational potential at short distances in the presence of compact extra dimensions~\cite{Zhou:2014xbw, Dahia:2015xxa,Dahia:2015bza}.

The comprehensive analysis of atomic hydrogen transitions from generalised interactions presented in this work is therefore well motivated from the perspective of physics beyond the Standard Model.

\clearpage

\section{Relativistic hydrogen atom}\label{sec:rha}
In order to introduce the method and notation used in this work, we begin by briefly reviewing the properties of the relativistic hydrogen atom, the wavefunction for which is found by solving the Dirac equation with a Coulomb potential
\begin{equation}\label{eq:diracEq}
    \left(i\slashed{\partial} - \gamma^0 \frac{Z\alpha_\mathrm{EM}}{r} - \mu\right) \psi_e(x) = 0,
\end{equation}
where $\slashed \partial = \gamma^\mu \partial_\mu$, with $\gamma^\mu$ one of the gamma matrices\footnote{We will work in the Dirac basis throughout, with $\gamma^{0} = \left(\begin{array}{cc}
    1 & 0 \\
    0 & -1
\end{array}\right)$, $\gamma^{i} = \left(\begin{array}{cc}
    0 & \sigma^i \\
    -\sigma^i & 0
\end{array}\right)$, and $\gamma^5 = \left(\begin{array}{cc}
    0 & 1 \\
    1 & 0
\end{array}\right)$, where $\sigma^i$ are the Pauli matrices.}, $Z$ is the atomic number of the nucleus, $\alpha_\mathrm{EM}$ is the fine-structure constant, $r$ is the radial coordinate and
\begin{equation}
    \mu = \frac{M m_e}{M + m_e} \simeq m_e,
\end{equation}
is the reduced electron mass, with $m_e$ and $M$ the electron and nuclear masses, respectively. To solve~\eqref{eq:diracEq} we follow the reasoning of~\cite{landauQED}, first noting that since the solution to the Dirac equation
\begin{equation}\label{eq:diracSolution}
    \mathcal{\psi}_e(x) = \left(\begin{array}{c}
         \phi(x)  \\
         \chi(x) 
    \end{array}\right),
\end{equation}
must be spherically symmetric, its components must be proportional to the spherically symmetric \textit{spinor spherical harmonics}, $\Omega_{j,l,m}(\theta, \phi)$, which are an extension of the regular spherical harmonics to the tensor product space of the orbital and spin angular momenta~\cite{Biedenharn_Louck_1984}. These satisfy the orthogonality relations
\begin{equation}\label{eq:jlmOrthogonality}
    \int d\cos\theta \,d\phi \, \Omega^{\dagger}_{j,l,m}(\theta,\phi) \,\Omega_{j',l',m'}(\theta,\phi) = \delta_{j,j'}\delta_{l,l'}\delta_{m,m'},
\end{equation}
with $j, l$, and $m$ the total, orbital and magnetic quantum numbers of the state, respectively. The solution~\eqref{eq:diracSolution} must also be an eigenstate of parity; under the action of the parity operator, the components of the Dirac bispinor transform as
\begin{equation}\label{eq:paritySpinor}
    \phi(t, \vec{x}) \xrightarrow[]{P} i\phi(t,-\vec{x}), \qquad \chi(t, \vec{x}) \xrightarrow[]{P} -i\chi(t,-\vec{x}),
\end{equation}
whilst the spinor spherical harmonics transform as
\begin{equation}\label{eq:parityHarmonic}
    \Omega_{j,l,m}(\vec{n}) \xrightarrow[]{P} \Omega_{j,l,m}(-\vec{n}) = (-1)^l \Omega_{j,l,m}(\vec{n}),
\end{equation}
where $\vec{n}$ is the unit vector along the direction defined by $\theta$ and $\phi$. From the combination of~\eqref{eq:paritySpinor} and~\eqref{eq:parityHarmonic}, it follows that in order for~\eqref{eq:diracSolution} to carry definite parity, the upper and lower components must be proportional to spinor spherical harmonics whose orbital angular momenta quantum numbers differ by one. Therefore, we introduce $\omega = 2j - l = l\pm1$, such that the solution to the Dirac equation for the relativistic hydrogen atom takes the form
\begin{equation}
    \psi_e(x) = \mathcal{U}_{n,j,l,m}(\vec{x}) e^{-iE_{n,j,l} t} = \left(\begin{array}{r}
         f_{n,j,l}(r) \Omega_{j,l,m}(\theta,\phi)  \\
         (-1)^{\frac{1}{2}(1+l-\omega)}g_{n,j,l}(r) \Omega_{j,\omega,m}(\theta,\phi) 
    \end{array}\right)e^{-iE_{n,j,l} t},
\end{equation}
where $n$ is the principal quantum number, $E_{n,j,l}$ is the energy of the state, and the phase factor on the lower component is chosen such that the radial wavefunctions $f$ and $g$ take a similar functional form. We give the full forms of both the radial and angular wavefunctions in Appendix~\ref{sec:wavefunction}.

When computing transition amplitudes we would like to use the orthogonality of the spinor spherical harmonics as much as possible, however, as we will see in Section~\ref{sec:rates}, transition amplitudes will frequently contain Pauli matrices acting on spinor spherical harmonics. Unfortunately, the action of Pauli matrices on spinor spherical harmonics in the $j,l,m$ basis is non-trivial, requiring a Clebsch-Gordan decomposition which significantly complicates the use of~\eqref{eq:jlmOrthogonality}. To overcome this issue, we follow~\cite{rose1961relativistic} and introduce the operator
\begin{equation}
    \hat{\kappa} = 1 + \vec{\sigma}\cdot \vec{L}, \qquad \hat{\kappa}\, \Omega_{\kappa,m} = -\kappa \,\Omega_{\kappa,m},
\end{equation}
with $\vec{L}$ the orbital angular momentum operator, and where
\begin{equation}\label{eq:kappaSeries}
    \kappa \in \{-1,+1,-2,+2,-3,+3,\dots\},
\end{equation}
corresponds one-to-one with the term symbol series
\begin{equation}
    L_J \in \left\{S_{\frac{1}{2}}, P_{\frac{1}{2}}, P_{\frac{3}{2}}, D_{\frac{3}{2}}, D_{\frac{5}{2}}, F_{\frac{5}{2}}, \dots\right\}.
\end{equation}
\begin{figure}[]
    \centering
    \includegraphics[scale=0.9]{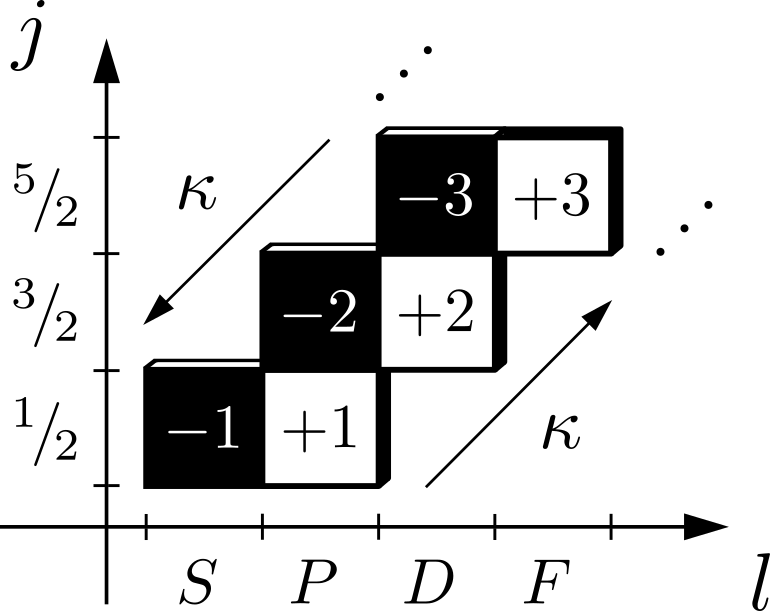}
    \caption{Correspondence between $\kappa$ and the hydrogen angular momentum states in spectroscopic notation.}
    \label{fig:kappaChessboard}
\end{figure}
That is, starting from $\kappa = -1$ with $S_{\frac{1}{2}}$ and moving along the series~\eqref{eq:kappaSeries}, we alternate between first incrementing $l$ and then $j$. This is illustrated in Figure~\ref{fig:kappaChessboard}, which we will later refer to when discussing allowed and forbidden transitions. More generally, we can relate $\kappa$, $j$ and $l$ using the relation
\begin{equation}\label{eq:kappaJL}
    \kappa = (l-j)(2j+1) \implies \kappa = \begin{cases}
        l, \quad & j = l - \frac{1}{2}, \\
        -l-1, \quad & j = l + \frac{1}{2},
    \end{cases} \qquad j = |\kappa| - \frac{1}{2}.
\end{equation}
The spinor spherical harmonics in the $\kappa,m$ basis satisfy analogous orthogonality relationships to~\eqref{eq:jlmOrthogonality}, and are related to those in the $j,l,m$ basis by
\begin{equation}
    \Omega_{\kappa,m}(\theta,\phi) = \Omega_{j,l,m}(\theta,\phi), \qquad i\Omega_{-\kappa,m}(\theta,\phi) = (-1)^{\frac{1}{2}(1 + l - \omega)}\Omega_{j,\omega,m}(\theta,\phi),
\end{equation}
allowing us to unify $j$ and $l$ and simplify the hydrogenic wavefunction to
\begin{equation}\label{eqref:hydrogenwavefunction}
    \mathcal{U}_{n,\kappa,m}(\vec{x}) = \left(\begin{array}{r}
         f_{n,\kappa}(r) \Omega_{\kappa,m}(\theta,\phi)  \\
         ig_{n,\kappa}(r) \Omega_{-\kappa,m}(\theta,\phi) 
    \end{array}\right).
\end{equation}
Importantly, the spinor spherical harmonics satisfy simple recurrence relations when acted on Pauli matrices. In particular, we will make extensive use of the relations~\cite{Szmytkowski:2007mzc}
\begin{align}
    \sigma^z \Omega_{\kappa,m}(\theta,\phi) &= -\frac{2m}{2\kappa +1}\Omega_{\kappa,m}(\theta,\phi) - \frac{2\sqrt{(\kappa+\frac{1}{2})^2 - m^2}}{|2\kappa + 1|}\Omega_{-\kappa-1, m}(\theta,\phi), \label{eq:spinorRecurrence1}\\
    \begin{split}
    \sigma^\pm \Omega_{\kappa,m}(\theta,\phi) &= \pm \sqrt{2}\frac{\sqrt{\kappa^2 - (m \pm \frac{1}{2})^2}}{2\kappa + 1}\Omega_{\kappa, m\pm 1}(\theta,\phi) \\
    & \qquad\qquad\qquad\qquad - \sqrt{2}\frac{\sqrt{(\kappa \pm m + \frac{1}{2})(\kappa \pm m + \frac{3}{2})}}{2\kappa + 1}\Omega_{-\kappa-1, m\pm 1}(\theta,\phi),\label{eq:spinorRecurrence2}
    \end{split}
\end{align}
which map spinor spherical harmonics back onto themselves under the action of the Pauli matrices, allowing us to always make use of the orthogonality relationship. Here we have also introduced $\sigma^{\pm} = \mp \frac{1}{\sqrt{2}} (\sigma^x \pm i\sigma^y)$, which belong to the spherical basis discussed in Appendix~\ref{sec:spherical-basis}.
%
%
\section{Transition rates}\label{sec:rates}
We will now work on deriving a general equation for the absorption and emission rates by a hydrogenic atom. Throughout, we will consider the absorption and emission of states with definite linear momentum and polarisation, rather than those of definite angular momentum. Starting with Fermi's golden rule, we have the total event rate
\begin{equation}\label{eq:goldenRule}
    d\Gamma = 2\pi |\mathcal{M}_{fi}|^2 \delta\left(\sum_i E_i - \sum_f E_f\right) d\rho,
\end{equation}
where $\mathcal{M}_{fi}$ denotes the transition amplitude, $E_i$ and $E_f$ denote the energies of the initial and final states, respectively, and where for an isotropic flux
\begin{equation}
    d\rho = \begin{cases}
        \displaystyle\,\,\prod_f \,\frac{d^3p_f}{(2\pi)^3} \,(1\pm f_f(p_f)), &\quad \mathrm{emission}, \\[14pt]
           \displaystyle\,\, \prod_i \,\frac{g_i d^3p_i}{(2\pi)^3}  \,|\vec{\beta}_i| f_i(p_i), &\quad \mathrm{absorption},
    \end{cases}
\end{equation}
with $g_i$ the number of degrees of freedom of each initial state species, and where $|\vec{\beta}_i| = |\vec{p}_i|/E_i$ and $f_i(p_i)$ are the velocity and distribution functions for species $i$, respectively. When considering emission processes, we take the upper sign for bosons which accounts for Bose enhancement, and the lower sign for fermions, which suppresses the process due to Pauli blocking. Notice that in~\eqref{eq:goldenRule} there is only a delta function fixing the energies, owing to the fact that we work with hydrogen states of definite angular momentum, which are not eigenstates of definite linear momentum. 

The last remaining component is the transition amplitude, which reads
\begin{equation}
    \mathcal{M}_{fi} = \bra{f}\int d^3x \mathcal{H}_\mathrm{int}(x)\ket{i},
\end{equation}
with $\ket{i}$ and $\bra{f}$ denoting the initial and final states, respectively. In all cases, the interaction Hamiltonian takes the generic form
\begin{equation}
    \mathcal{H}_\mathrm{int}(x) = \sum_L \left(\bar\psi_e(x) \Gamma^{\{\mu\}}_L \psi_e(x)\right) \mathcal{O}_{L,\{\mu\}}(x), \qquad \Gamma_L^{\{\mu\}} \in \{1,\gamma^5, \gamma^\mu, \gamma^\mu\gamma^5, \sigma^{\mu\nu}\},
\end{equation}
where $L \in \{S, P, V, A, T\}$ denotes one of the five linearly independent Lorentz structures, corresponding to scalar, pseudoscalar, vector, axial-vector and tensor, respectively. We have also defined $\sigma^{\mu\nu} = \frac{i}{2}[\gamma^\mu,\gamma^\nu]$, and used $\mathcal{O}_L$ to denote the interaction operator with a set of Lorentz indices $\{\mu\}$, composed of field operators for the particles coupled to the hydrogen electrons and couplings, which may be dimensionful. Each of the five Lorentz structures gives rise to a different set of transitions, which are tabulated in Table~\ref{tab:lorentzStructures} alongside their equivalent electromagnetic multipole and corresponding ``move'' in $\kappa$-space as illustrated in Figure~\ref{fig:kappaChessboard}. 
\begin{table}[]
\centering
\renewcommand{\arraystretch}{2}
\setlength\tabcolsep{5mm}
\begin{tabular}{Sc|Sc|Sc|Sc}
$\kappa\to\ldots$ & Transitions & Couplings & Multipole \\ \hline\hline
\raisebox{1mm}{$\kappa$} & \includegraphics[]{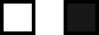} & S, A, T & E0, M1, E2, \dots \\ \hline
\raisebox{2.5mm}{$-\kappa$} & \hspace{1mm}\raisebox{-1mm}{\includegraphics[]{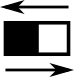}}&  \raisebox{2mm}{P, V, A, T} & \raisebox{2mm}{M0, E1, M2, \dots} \\ \hline
\raisebox{2.5mm}{$-\kappa-1$} & \raisebox{-1mm}{\includegraphics[]{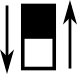}} & \raisebox{2mm}{A, T} & \raisebox{2mm}{M1, E2, M3, \dots} \\ \hline
\raisebox{3.5mm}{$\kappa+1$} & \hspace{-2mm}\raisebox{-1mm}{\includegraphics[]{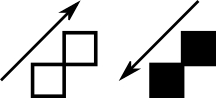}} & \raisebox{3mm}{V, T} & \raisebox{3mm}{E1, M2, E3, \dots} \\ \hline
\raisebox{3.5mm}{$\kappa-1$} & \hspace{-2mm}\raisebox{-1mm}{\includegraphics[]{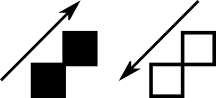}} & \raisebox{3mm}{V, T} & \raisebox{3mm}{E1, M2, E3, \dots} \\ \hline
\raisebox{3.5mm}{$-\kappa+1$} & \raisebox{-1mm}{\includegraphics[]{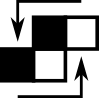}} & \raisebox{3mm}{A, T} & \raisebox{3mm}{M1, E2, M3, \dots}
\end{tabular}
\caption{Allowed transitions for each Lorentz structure, alongside their corresponding move in $\kappa$-space represented in Figure~\ref{fig:kappaChessboard}, and their equivalent electromagnetic multiple moment.}
\label{tab:lorentzStructures}
\end{table}

The hydrogen field operator is 
\begin{equation}
    \psi_e(x) = \sum_{n,\kappa,m} \left(a_{n,\kappa,m}\, \mathcal{U}_{n,\kappa,m}(x) + b^\dagger_{n,\kappa,m} \mathcal{V}^\dagger_{n,\kappa,m}(x)\right), 
\end{equation}
where $\mathcal{U}_{n,\kappa,m}$ is the hydrogen wavefunction given in \eqref{eqref:hydrogenwavefunction}, and $\mathcal{V}_{n,\kappa,m}$ is the corresponding antiparticle solution, which we do not specify explicitly here but which can be found by charge conjugation. The creation and annihilation operators satisfy the discrete anticommutation relations
\begin{equation}
    \left\{a_{n,\kappa,m}, a^\dagger_{n',\kappa',m'}\right\} = \left\{b_{n,\kappa,m}, b^\dagger_{n',\kappa',m'}\right\} = \delta_{n,n'} \delta_{\kappa, \kappa'} \delta_{m,m'}.
\end{equation}
To find the amplitude, the anticommutators above are to be coupled with the definitions of the external states
\begin{equation}
    \ket{i} = a^\dagger_{n,\kappa,m}\ket{i_\mathcal{O}}, \qquad \bra{f} = \bra{f_\mathcal{O}} a_{n',\kappa',m'},
\end{equation}
and 
with $\ket{i_\mathcal{O}}$ and $\bra{f_\mathcal{O}}$ denoting similarly defined states for the absorbed or emitted particles, respectively. As such, our amplitude will read
\begin{equation}\label{eq:ampPhasefactor}
    \begin{split}
    \mathcal{M}_{fi} &= \sum_{L}\int d^3x \left( \bar{\mathcal{U}}_{n',\kappa',m'}(x)\Gamma^{\{\mu\}}_L\mathcal{U}_{n,\kappa,m}(x)\right) \bra{f_\mathcal{O}}\mathcal{O}_{L,\{\mu\}}(x)\ket{i_\mathcal{O}} \\
    &=\sum_L\langle\mathcal{O}_{L,\{\mu\}}\rangle\int d^3x\, \left( \bar{\mathcal{U}}_{n',\kappa',m'}(x)\Gamma_L^{\{\mu\}}\mathcal{U}_{n,\kappa,m}(x)\right) e^{-i\vec{k}\cdot \vec{x}},
    \end{split}
\end{equation}
in which $\langle\mathcal{O}_{L,\{\mu\}}\rangle$ is defined as the expectation value with the phase factored out, and where $\vec{k}$ is some function of the momenta of the absorbed or emitted particles. The presence of the phase factor in~\eqref{eq:ampPhasefactor} significantly complicates the evaluation of these amplitudes, however, as we show in Appendix~\ref{sec:dipoleApproximation}, the dipole approximation $(e^{i\vec{k}\cdot \vec{x}} \simeq 1)$ is sufficiently precise in most cases. After squaring, we are left with
\begin{equation}\label{eq:ampSq}
    \begin{split}
    |\mathcal{M}_{fi}|^2 &= \sum_{L,L}\langle\mathcal{O}_{L,\{\mu\}}\rangle\,\langle\mathcal{O}_{L',\{\nu\}}\rangle^*\, A_{L,L'}^{\{\mu\}\{\nu\}},
    \end{split}
\end{equation}
where we have defined the \textit{atomic tensor}
\begin{equation}\label{eq:atomicTensor}
    A_{L,L'}^{\{\mu\}\{\nu\}} = \left(\int d^3x\,  \bar{\mathcal{U}}_{n',\kappa',m'}(x)\Gamma^{\{\mu\}}_L\mathcal{U}_{n,\kappa,m}(x)\right)\left(\int d^3x'\,  \bar{\mathcal{U}}_{n',\kappa',m'}(x')\Gamma^{\{\nu\}}_{L'}\mathcal{U}_{n,\kappa,m}(x')\right)^*.
\end{equation}
The components of the atomic tensor are computed by making extensive use of the spinor spherical harmonic identities presented in Section~\ref{sec:rha}, and then performing the resulting radial integrals numerically. Whilst not conceptually challenging, computing these amplitudes manually can take a significant amount of time, in particular when interference terms between different Lorentz structures are present, or if the amplitude needs to be evaluated for a wide range of quantum numbers. Additionally, in cases where high numerical precision is required, care must be taken when evaluating the radial integrals, which feature rapidly diverging hypergeometric functions attenuated by an exponential function with a large negative exponent. This is particularly challenging at large values of the principal quantum number, $n$. For these reasons, we have developed \texttt{CINCO}, a lightweight program for rapidly constructing amplitudes of the form~\eqref{eq:ampSq} and computing the associated radial integrals to high precision.
\subsection{Automated transition amplitudes with \texttt{CINCO}}\label{sec:cinco}
The object at the heart of the hydrogen transition amplitudes is the atomic tensor~\eqref{eq:atomicTensor}, which by rewriting the gamma matrices in terms of Pauli matrices, and using the orthogonality and recurrence relations for the spherical spinor harmonics can always be reduced to a product of the form (with Lorentz indices suppressed on the RHS)
\begin{equation}
    A_{L,L'}^{\{\mu\}\{\nu\}} = \left(\sum_t C_{\kappa,m}^{(t)} \delta_{t} \mathcal{I}_t\right)\left(\sum_{t'} C_{\kappa,m}^{(t')*} \delta_{t'} \mathcal{I}_{t'}\right),
\end{equation}
where the summation runs over the transitions induced by each vertex, $\delta$ represents a pair of Kronecker delta functions relating $\kappa$ to  $\kappa'$ and $m$ to $m'$, which serves as a selection rule, $C_{\kappa,m}$ is some complex valued coefficient depending only on $\kappa$ and $m$, and $\mathcal{I}$ is some linear combination of the four possible, real valued, radial integrals
\begin{equation}\label{eq:radialIntegrals}
    \begin{alignedat}{2}
    \mathcal{I}_{ff} &= \int dr\, r^2 f_{n',\kappa'}(r) f_{n,\kappa}(r) \sim Z^{0}, \qquad \mathcal{I}_{fg} &&= \int dr\, r^2 f_{n',\kappa'}(r) g_{n,\kappa}(r) \sim Z^{1}, \\
    \mathcal{I}_{gf} &= \int dr\, r^2 g_{n',\kappa'}(r) f_{n,\kappa}(r) \sim Z^{1}, \qquad \mathcal{I}_{gg} &&= \int dr\, r^2 g_{n',\kappa'}(r) g_{n,\kappa}(r) \sim Z^{2},
    \end{alignedat} 
\end{equation} 
each of which scales with a different power of $Z$, which we derive in Appendix~\ref{sec:wavefunction}. 
For example, if we considered the $\gamma^z$ vertex, we would have
\begin{equation}
    \begin{split}
    \sum_t C_{\kappa,m}^{(t)} \delta_{t} \mathcal{I}_t &= 2im \delta_{\kappa',-\kappa} \delta_{m',m} \left(\frac{\mathcal{I}_{fg}}{2\kappa-1} + \frac{\mathcal{I}_{gf}}{2\kappa+1}\right)+ \frac{2i\sqrt{(\kappa+\frac{1}{2})^2-m^2}}{|2\kappa+1|} \delta_{\kappa',\kappa+1} \delta_{m',m} \mathcal{I}_{gf} \\
    &- \frac{2i\sqrt{(\kappa-\frac{1}{2})^2-m^2}}{|2\kappa-1|} \delta_{\kappa',\kappa-1} \delta_{m',m} \mathcal{I}_{fg}.
    \end{split}
\end{equation}
We give the form of $C_{\kappa,m}$, $\delta$ and $\mathcal{I}$ for the complete set of Lorentz structures and indices in Appendix~\ref{sec:transitionTables}, working in the spherical basis, which is discussed in Appendix~\ref{sec:spherical-basis}. This is the foundation of \texttt{CINCO}; given an amplitude structure (\textit{e.g.} ``V+A'', ``P+T'') and a set of quantum numbers for the initial and final states, \texttt{CINCO} uses the entries of Tables~\ref{tab:transitions-scalar},~\ref{tab:transitions-vector} and~\ref{tab:transitions-tensor} to construct the elements of the atomic tensor, which are then simplified to remove orthogonal and forbidden terms using the \texttt{SymPy} package~\cite{sympy}. The elements of the atomic tensor are then multiplied by the corresponding expectation values $\left\langle\mathcal{O}\right\rangle$ to give the squared amplitude~\eqref{eq:ampSq}, the \LaTeX\ for which is then printed to a file after transforming back to the Cartesian basis so that the usual relations for spin and polarisation sums can be used if required. \texttt{CINCO} also computes the four radial integrals~\eqref{eq:radialIntegrals} and the transition energy $\Delta E_{fi}$ to high precision by making use of the \texttt{mpmath} package~\cite{mpmath}, the values of which are saved to a separate file. If specified at runtime, \texttt{CINCO} will instead compute the $m$-averaged amplitude
\begin{equation}\label{eq:ampSqAveraged}    \left\langle|\mathcal{M}_{fi}|^2\right\rangle_{m,m'} = \frac{1}{2|\kappa|}\sum_{m,m'}\sum_{L,L}\langle\mathcal{O}_{L,\{\mu\}}\rangle\,\langle\mathcal{O}_{L',\{\nu\}}\rangle^*\, A_{L,L'}^{\{\mu\}\{\nu\}},
\end{equation}
which is preferable in situations where the initial and final state polarisations cannot be controlled. As discussed in Section~\ref{sec:introduction}, it may also be interesting to explore the effect of modified fundamental constants on transition amplitudes. For this reason we also include the ability to specify fractional modifications to $\alpha_\mathrm{EM}$ and $m_e$ at runtime. We give examples of amplitudes computed using \texttt{CINCO} in Section~\ref{sec:examples}, whilst the code and documentation for \texttt{CINCO} can be found at \href{https://gitlab.com/JShergold/cinco}{gitlab.com/JShergold/cinco}. 
%
%
\subsection{Enhanced transition rates for large $Z$}
It is important to emphasise the impact of the $Z$-scaling for different atomic transitions in hydrogen. We first observe that the different radial integrals in \eqref{eq:radialIntegrals} scale at most like $Z^2$. The amplitude squared for any transition can then obtain at most a contribution proportional to $\mathcal{I}_{gg}^2 \sim \, Z^4$ from the radial integrals. The second observation is that the transition energy gap scales as $\Delta E_{fi} \sim\,\alpha_\mathrm{EM}^2 Z^2$. 
As a result, non-renormalisable effective operators of the type
\begin{equation}\label{eq:effop}
    \mathcal{H}_\mathrm{int}(x) = \sum_L \frac{1}{\Lambda^{d-4}}\left(\bar\psi_e(x) \Gamma^{\{\mu\}}_L \psi_e(x)\right) \mathcal{O}_{L,\{\mu\}}(x),
\end{equation}
where $\Lambda$ is the suppression scale and $d$ is the combined mass dimension of $\mathcal{O}$ and the hydrogen electron bilinear, induce transition rates that can scale as strongly with $Z$ as
\begin{align}\label{eq:Zscaling}
\Gamma \sim \Delta E_{fi}\left(\frac{\Delta E_{fi}} {\Lambda}\right)^{d-4} \mathcal{I}_{gg}^2\sim (\alpha_\mathrm{EM} Z)^{2(d-1)}\,.
\end{align}
The exact strength of this remarkable power-law scaling depends on the structure of the underlying theory responsible for the operator in \eqref{eq:effop}, which dictates the different mass scales that can appear in the numerator of \eqref{eq:Zscaling}, and on the specific transition that determines the dominant radial integral in \eqref{eq:radialIntegrals}. For example, the absorption and emission rate of neutrino-antineutrino pairs in stellar hydrogen scales as $\Gamma_{\nu\bar\nu}\sim (\alpha_\mathrm{EM} Z)^{14} G_F^2$~\cite{PhysRev.151.1189}. 

Transitions in highly charged hydrogenic atoms can therefore have enhanced sensitivity to new physics effects. This scaling is even more pronounced than the $Z$-scaling for the binding energy ($\sim Z^2$), the hyperfine splitting ($\sim Z^3$), QED effects ($\sim Z^4$) and the Stark effect ($\sim Z^6$) in hydrogenic ions. Future searches for absorptions with ion clocks using highly charged ions could therefore be uniquely sensitive to light new physics. Probes with different ions can be utilised to distinguish the Lorentz structure of the interaction and the dimension of the operator by probing the $Z$-scaling for different transitions. We stress that this is an independent motivation to use future highly charged ion clocks in searches for new physics, extending the physics case presented in~\cite{Kozlov:2018mbp}.
\subsection{New long-range forces with \texttt{CINCO}}
We now briefly comment on the possibility of constraining new, long-range forces with \texttt{CINCO}. Suppose that we have two long-range forces contributing to the central potential, electromagnetism, with coupling $\beta$, and some new force with coupling strength $\eta$ and some alternative radial scaling $\mathcal{R}(r)$, which in general will take the form
\begin{equation}
    \mathcal{R}(r) = \frac{1}{r^n} e^{-mr},
\end{equation}
with $m$ the mass of the species responsible, and $n \geq 1$. The modified potential will then take the form, assuming a single proton,
\begin{equation}\label{eq:newPotential}
    U(r) = \frac{\beta}{r} + \eta \mathcal{R}(r).
\end{equation}
Our solutions to the Dirac equation work under the ansatz of a pure $\frac{1}{r}$ potential, with coupling constant $\alpha$, that we set to the fine-structure constant. There are then two situations that can arise.
\begin{enumerate}
    \item $\mathcal{R}(r) \neq \frac{1}{r}$: In this scenario, the measured distribution of energy levels will deviate from the predictions of a purely $\frac{1}{r}$ potential. This can alternatively be thought of as having a $\frac{1}{r}$ potential and coupling constant that depends on radius, effectively satisfying
    \begin{equation}
        \frac{\alpha(r)
        }{r} = \frac{\beta}{r} + \eta \mathcal{R}(r).
    \end{equation}
    Within \texttt{CINCO}, one would then notice that for fixed $\alpha$, our results will agree for some values of $r$, or equivalently for some orbitals, and begin to deviate as we move away from these orbitals. To recover the correct results, one could then adjust the value of $\alpha$ within \texttt{CINCO} in line with the measurements and build the profile of $\alpha(r)$. From here, $\beta$ can be recovered by fitting $\alpha(r)$ at large radii where $\mathcal{R}(r) \to 0$, and the remaining parameters by fitting the remainder of $\alpha(r)$. As discussed in~\cite{Delaunay:2022grr}, the measured value of the fine-structure constant includes the effects of new physics, often leading to inconsistencies when attempting to constrain new forces. By isolating $\beta$, the \textit{true} electromagnetic coupling, at large radii, this method allows us to consistently distinguish the contributions from new forces.
    \item $\mathcal{R}(r) = \frac{1}{r}$: In this scenario, as pointed out in~\cite{Delaunay:2022grr}, distinguishing between $\beta$ and $\eta$ for purely proton-philic interactions becomes impossible as
    \begin{equation}
        \alpha = \beta + \eta.
    \end{equation}
    Here, the measured value of the fine-structure constant as extracted from atomic transitions would be exactly equal to the sum of the electromagnetic and new physics couplings. This degeneracy could be lifted, however, if the new mediator couples to species other than protons. For example, if the new species also couples to neutrons, measurements along the isotopic chain could be used to isolate the neutron insensitive $\beta$, and in turn fit $\eta$.
\end{enumerate}
It is also possible to use a similar method to constrain neutron-philic interactions, or more generally, couplings to species other than protons \textit{e.g.} muonic hydrogen, positronium. Focusing on the neutron-philic case, and working with the ansatz $\eta = \eta(N)$, and $\alpha = \alpha(r,N)$, with $N$ the number of neutrons, one can compare experimental data with that of \texttt{CINCO} at a fixed radius, or equivalently fixed energy shell. The value of $\alpha$ within \texttt{CINCO} can then be varied to build up a profile of the effective coupling required to reproduce the data for each $N$, and fit $\eta(N)$. Unlike new, exclusively proton-philic forces, this should also be possible when $\mathcal{R}(r) = \frac{1}{r}$. 

In future works, we plan to include the effects of new forces in our hydrogen atom solutions, including neutron-philic interactions, which will expedite this process significantly.

\section{Examples}\label{sec:examples}
We now give some concrete examples of amplitudes computed with \texttt{CINCO} to demonstrate its capabilities. We will start with what is possibly the simplest transition, the photon $2P_{3/2} \to 1S_{1/2}$ de-excitation, and then move onto a more complicated example for dark matter. 
\subsection{Photon $2P_{3/2} \to 1S_{1/2}$ transition}\label{sec:photonRate}
We begin with the electron-photon interaction Hamiltonian, which reads
\begin{equation}
    \mathcal{H}_\mathrm{int}(x) = e A_\mu(x)\bar\psi_e(x) \gamma^\mu \psi_e(x) + \mathcal{O}\left(\frac{\Delta E_{fi}}{m_e}\right),
\end{equation}
where $e$ is the electron charge, and the neglected terms are those from the tensor operators giving rise to the electron magnetic and electric dipole moments, which are approximately suppressed by the ratio of the transition energy, $\Delta E_{fi}$, to the electron mass, $m_e$. We note, however, that these are the sole surviving terms for transitions that cannot proceed through the vector vertex. A full list of these can be found in Tables~\ref{tab:transitions-vector} and~\ref{tab:transitions-tensor}. 

For consistency with our definition of the hydrogen field operator, we expand our photon field operator as
\begin{equation}
    A_\mu(x) = \sum_{\vec{k},h} \left(c_{\vec{k},h}\psi_{A,\mu} e^{i\vec{k}\cdot \vec{x}} + c_{\vec{k},h}^\dagger\psi_{A,\mu}^* e^{-i\vec{k}\cdot \vec{x}}\right),
\end{equation}
where the sum runs over the photon momentum and helicity, and the creation and annihilation operators satisfy the commutation relation
\begin{equation}
    \left[c_{\vec{p},h}, c^\dagger_{\vec{q},h'}\right] = \delta_{\vec{p},\vec{q}}\,\delta_{h,h'},
\end{equation}
with all other commutators zero. These results are equivalent to those in the continuum limit, which can be restored with the replacements $\sum_{\vec{k}} \to \int \frac{d^3k}{(2\pi)^3}$ and $\delta_{\vec{p},\vec{q}} \to (2\pi)^3 \delta^{(3)}(\vec{p} - \vec{q})$. The photon wavefunction, normalised to one per unit volume is given by\footnote{For completeness we also give the appropriately normalised wavefunctions for spin-0 and spin-$\frac{1}{2}$ particles, which are $\psi_\phi = \frac{1}{\sqrt{2E_\phi}}$ and $\psi_\chi = \frac{1}{\sqrt{2E_\chi}} u_h$, respectively, where $u_h$ is a Dirac spinor with helicity $h$.}
\begin{equation}
    \psi_{A,\mu} = \frac{1}{\sqrt{2E_\gamma}}\epsilon_{h,\mu},
\end{equation}
with $\epsilon_{h,\mu}$ the photon polarisation vector and $E_\gamma = |\vec{k}|$. With these definitions, we can immediately write down our expectation value
\begin{equation}
    \left\langle\mathcal{O}_{V,\mu}\right\rangle = e\bra{\gamma}A_\mu\ket{0} = \frac{e}{\sqrt{2E_\gamma}} \epsilon_{h,\mu}^*,
\end{equation}
which appears in the amplitude
\begin{equation}
    \begin{split}
    \mathcal{M}_{fi} &= \left\langle\mathcal{O}_{V,\mu}\right\rangle \int d^3x\, \big( \bar{\mathcal{U}}_{n',\kappa',m'}(x)\gamma^{\mu}\mathcal{U}_{n,\kappa,m}(x)\big),
    \end{split}
\end{equation}
\begin{figure}[]
    \centering
    \begin{subfigure}{0.49\textwidth}
    \includegraphics[width=\textwidth]{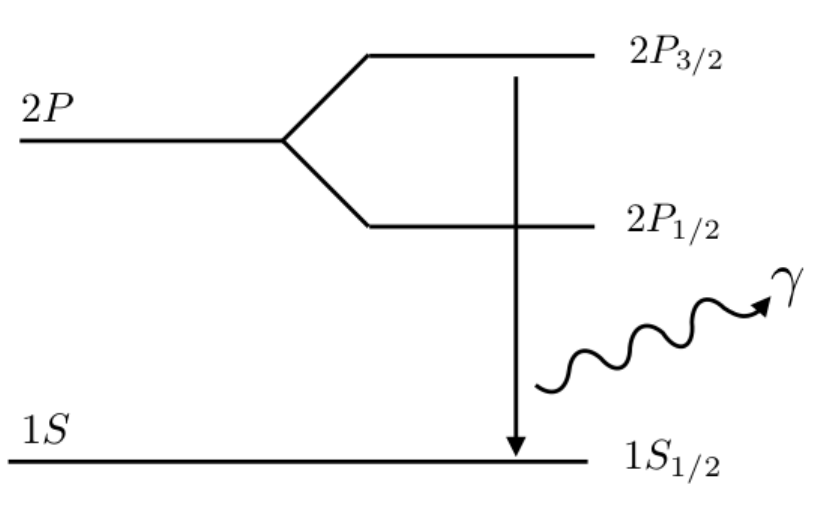}
    \caption{}
    \label{fig:second}
\end{subfigure}
    \begin{subfigure}{0.49\textwidth}
    \includegraphics[width=\textwidth]{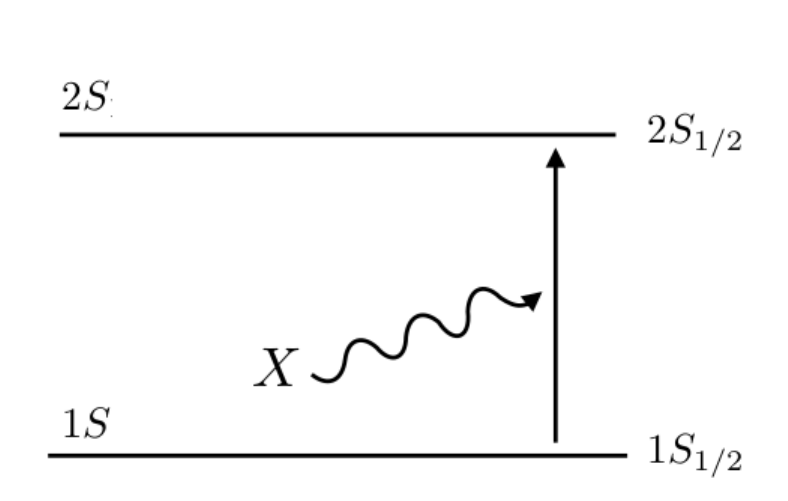}
    \caption{}
    \label{fig:second}
\end{subfigure}
    \caption{a) Hydrogen energy levels for the $2P_{3/2} \to 1S_{1/2}$ de-excitation discussed in Section~\ref{sec:photonRate}, and b) energy levels for the absorption of a spin-1 dark matter particle with axial-vector couplings as discussed in Section~\ref{sec:darkvector}.}
    \label{fig:examples}
\end{figure}
with $n = 2$, $n' = 1$, $\kappa = -2$, and $\kappa = -1$. With the amplitude written in this form, we use \texttt{CINCO} in amplitude mode to immediately arrive at the squared amplitude, averaged over $m$ and summed over $m'$
\begin{equation}
\begin{split}\left\langle|\mathcal{M}_{fi}|^2\right\rangle_{m,m'} &= \frac{4}{9} \left(\langle\vec{\mathcal{O}}_V\rangle \cdot \langle\vec{ \mathcal{O}}_V\rangle^*\right) \mathcal{I}_{gf}^{2},
    \end{split}
\end{equation}
with the Cartesian inner product
\begin{equation}
   \left(\langle\vec{\mathcal{O}}_V\rangle \cdot \langle\vec{\mathcal{O}}_V\rangle^*\right) = \langle{\mathcal{O}}_{V,x}\rangle\langle\mathcal{O}_{V,x}\rangle^* + \langle\mathcal{O}_{V,y}\rangle\langle\mathcal{O}_{V,y}\rangle^*+\langle\mathcal{O}_{V,z}\rangle\langle\mathcal{O}_{V,z}\rangle^*,
\end{equation}
and where the radial integral and transition energy take the values
\begin{equation}
    \mathcal{I}_{gf} = -0.0017654, \qquad \Delta E_{fi} = 10.199\,\mathrm{eV},
\end{equation}
respectively. Substituting in our current, this becomes
\begin{equation}
    \begin{split}
    \left\langle|\mathcal{M}_{fi}|^2\right\rangle_{m,m'} &= \frac{8\pi \alpha_\mathrm{EM}}{9E_\gamma} \left(\vec{\epsilon}_h \cdot \vec{\epsilon}_h^*\right)  \mathcal{I}_{gf}^{2}.
    \end{split}
\end{equation}
In the final step, we sum over the two physical helicities of the final state photon and use the identity
\begin{equation}
    \sum_{h} \epsilon_\mu \epsilon_\nu^* = - g_{\mu\nu} + \frac{k_\mu \bar k_\nu + \bar k_\mu k_\nu}{k\cdot \bar k} \implies \sum_h \left(\vec{\epsilon}_h \cdot \vec{\epsilon}_h^*\right) = 2,
\end{equation}
with $\bar k_0 = k_0$ and $\bar k_i = -k_i$, to find
\begin{equation}
    \begin{split}
    \left\langle|\mathcal{M}_{fi}|^2\right\rangle &= \frac{16\pi \alpha_\mathrm{EM}}{9E_\gamma}  \mathcal{I}_{gf}^{2}.
    \end{split}
\end{equation}
Finally, to find the transition rate we substitute the amplitude into Fermi's golden rule and perform the integral over the final state photon momenta, leaving us with
\begin{equation}\label{eq:photonRate}
    \begin{split}
    \Gamma_{2S_{3/2}\to 1S_{1/2}} &= \frac{16\alpha_\mathrm{EM}}{9} \Delta E_{fi}\mathcal{I}_{gf}^{2} \\
    &= 6.2650 \cdot 10^{8}\,\mathrm{s}^{-1},
    \end{split}
\end{equation}
which agrees with the value $\Gamma_{2S_{3/2}\to 1S_{1/2}} = 6.2647\cdot 10^{8}\,\mathrm{s}^{-1}$ given by NIST~\cite{NIST_ASD} to within $0.005\%$. The small discrepancy between our value and that of NIST is due to the terms beyond the dipole approximation, here dominated by the M2 term, which contributes at the $\mathcal{O}\left(10^{-5}\right)$ level~\cite{10.1063/1.1796671}.
%
%
\subsection{Dark vector $1S_{1/2} \to 2S_{1/2}$ transition}\label{sec:darkvector}
We now turn our attention to a simple dark matter model with a massive, leptophilic vector boson $X$ with interaction Hamiltonian
\begin{equation}
    \mathcal{H}_\mathrm{int}(x) = X_\mu(x)\bar\psi_e(x) \gamma^\mu (g_L P_L + g_R P_R) \psi_e(x),
\end{equation}
with $g_L$ and $g_R$ the left and right chiral couplings, and $P_L$ and $P_R$ the corresponding projection operators. The addition of the axial couplings allows for transitions otherwise forbidden to photons, with the exception of through the heavily suppressed magnetic and electric dipole operators. We focus on the hydrogen $1S_{1/2} \to 2S_{1/2}$ transition, which could be used to search for dark matter via its absorption by hydrogen.

As before, we first compute the vector and axial-vector expectation values
\begin{equation}
    \langle\mathcal{O}_{V,\mu}\rangle = \frac{g_R + g_L}{2} \frac{1}{\sqrt{2E_X}} \epsilon_\mu, \qquad \langle\mathcal{O}_{A,\mu}\rangle = \frac{g_R - g_L}{2} \frac{1}{\sqrt{2E_X}} \epsilon_\mu,
\end{equation}
and then use \texttt{CINCO} to compute the $m$-averaged squared amplitude,
\begin{equation}
    \begin{split}
        \left\langle|\mathcal{M}_{fi}|^2\right\rangle_{m,m'} &= \left(\langle\vec{\mathcal{O}}_A\rangle \cdot \langle\vec{\mathcal{O}}_A\rangle^*\right) \left(\mathcal{I}_{ff}^{2} - \frac{2}{3} \mathcal{I}_{ff} \mathcal{I}_{gg} + \frac{\mathcal{I}_{gg}^{2}}{9}\right) \\
        &+  \big|\langle{\mathcal{O}_{A,0}}\rangle\big|^{2} \left(2 \mathcal{I}_{fg}^{2} - 4 \mathcal{I}_{fg} \mathcal{I}_{gf} + 2 \mathcal{I}_{gf}^{2}\right)\\
        &+ \operatorname{Im}{\Big(\langle\mathcal{O}_{A,0}\rangle\langle\mathcal{O}_{A,x}\rangle^*\Big)}\left( 2 \mathcal{I}_{ff} \mathcal{I}_{gf} - 2 \mathcal{I}_{ff} \mathcal{I}_{fg} + \frac{2}{3} \mathcal{I}_{fg} \mathcal{I}_{gg} - \frac{2}{3} \mathcal{I}_{gf} \mathcal{I}_{gg}\right) 
    \end{split}
\end{equation}
which has no dependence on the vector operator, as expected. \texttt{CINCO} also gives the values of the radial integrals
\begin{equation}
    \begin{alignedat}{3}
    \mathcal{I}_{ff} &= -5.5785\cdot 10^{-6}, \qquad &\mathcal{I}_{fg} &= 2.0354\cdot 10^{-8}, \\
    \mathcal{I}_{gf} &=-0.0015289, \qquad &\mathcal{I}_{gg} &= 5.5785\cdot 10^{-6},
    \end{alignedat}
\end{equation}
and the transition energy
\begin{equation}
    \Delta E_{fi} = 10.199\,\mathrm{eV}.
\end{equation}
Next, we sum over the three physical polarisations of the vector boson, using the identity
\begin{equation}
    \sum_h \epsilon_{h,\mu} \epsilon^*_{h,\nu} = -g_{\mu\nu} + \frac{k_\mu k_\nu}{m_X^2},
\end{equation}
which allows us to replace
\begin{equation}
    \sum_h \left(\vec{\epsilon}_h\cdot\vec{\epsilon}_h^*\right) = 3 + \frac{|\vec{k}|^2}{m_X^2}, \qquad \sum_h \left|\epsilon_{h,0}\right|^{2}= -1 + \frac{E_X^2}{m_X^2}, \qquad \sum_h \operatorname{Im}{\left(\epsilon_{h,x} \epsilon_{h,0}^*\right)}= 0,
\end{equation}
where the last equality follows from the imaginary part operator commuting with the summation.
Substituting into the golden rule and integrating over all dark matter momenta in the galactic reference frame\footnote{We integrate over galaxy frame momenta to keep the integration limits simple. In exchange, we must integrate over a more complicated distribution function. See Appendix A of~\cite{Rostagni:2023eic} for more details.}, $\vec{k}_\mathrm{gal}$, we are left with the absorption rate per hydrogen atom
\begin{equation}
    \begin{split}
    \Gamma_{1S_{1/2}\to 2S_{1/2}} = \frac{3(g_L - g_R)^2}{32\pi^2}\frac{1}{m_X^2}\int d^3k_\mathrm{gal} \frac{|\vec{k}|}{E_X^2} \big[&C_1 (E_X^2 + 2 m_X^2) + C_2 |\vec{k}|^2 \big] \\
    &\times f\left(\vec{k}_\mathrm{gal} - m_X \vec{\beta}_\Earth\right)\delta(E_X - \Delta E_{fi}),
    \end{split}
\end{equation}
where $C_1 = 5.5324\cdot 10^{-11} \sim Z^4$ and $C_2 = 4.6752\cdot 10^{-6} \sim Z^2$ are composed of radial integrals, and $\vec{k}_\mathrm{gal}$ satisfies
\begin{equation}
    \vec{k} \simeq \vec{k}_\mathrm{gal} + m_X \vec{\beta}_\Earth = |\vec{k}_\mathrm{gal}|\left(\begin{array}{c}
         \cos\phi \sin\theta  \\
         \sin\phi \sin\theta \\
         \cos\theta
    \end{array}\right) + m_X |\vec{\beta}_\Earth|\left(\begin{array}{c}
         0  \\
         0 \\
         1
    \end{array}\right),
\end{equation}
with $|\vec{\beta}_\Earth| \simeq 10^{-3}$ the relative frame velocity. To perform the integral we follow~\cite{Rostagni:2023eic} and assume an isothermal spherical halo, such that in the lab frame the dark matter is distributed according to 
\begin{equation}
    f\left(\vec{k}_\mathrm{gal} - m_X \vec{\beta}_\Earth\right) = \frac{\rho_\mathrm{DM}}{m_X}\left(\frac{2\pi}{m_X^2 \sigma^2}\right)^\frac{3}{2}e^{-\frac{|\vec{k}_\mathrm{gal}|^2 + m_X^2 |\vec{\beta}_\Earth|^2}{2m_X^2 \sigma^2}}e^{\frac{|\vec{k}_\mathrm{gal}||\vec{\beta}_\Earth|\cos\theta}{m_X \sigma^2}},
\end{equation}
with $\sigma \simeq \sqrt{2}|\vec{\beta}_\Earth|$ the velocity dispersion and $\rho_\mathrm{DM} \simeq 0.4\,\mathrm{GeV}\,\mathrm{cm}^{-3}$ is the local dark matter density~\cite{Read:2014qva}. This gives the absorption rate to leading order in $|\vec{\beta}_\Earth|$
\begin{equation}
    \begin{split}
    \Gamma_{1S_{1/2}\to 2S_{1/2}} \simeq \frac{3(g_L - g_R)^2 \sqrt{\pi}}{2\sqrt{2}|\vec{\beta}_\Earth|^2}\frac{\rho_\mathrm{DM}\sqrt{m_{X}T_{X}}}{\Delta E_{fi}^2
   m_{X}^4} \big[C_1 \big(&\Delta E_{fi}^2+2 m_{X}^2\big)+ 2C_2 m_{X} T_{X}\big] \\
   &\times\sinh \left(\frac{4}{|\vec{\beta}_\Earth|}\sqrt{\frac{2T_X}{m_X}}\right)e^{-\frac{2 T_{X}}{|\vec{\beta}_\Earth|^2 m_{X}}},
   \end{split}
\end{equation}
where $T_X = \Delta E_{fi} - m_X \ll m_X$ is the kinetic energy of the absorbed vector boson. As such, the term proportional to $C_2$ contributes approximately $1\%$ to the total rate, despite the fact that $C_2 \gg C_1$, and we may write
\begin{equation}
    \begin{split}
    \Gamma_{1S_{1/2}\to 2S_{1/2}} &\simeq \frac{9(g_L - g_R)^2 C_1 \sqrt{\pi}}{2\sqrt{2}|\vec{\beta}_\Earth|^2}\frac{\rho_\mathrm{DM}\sqrt{T_{X}}}{\Delta E_{fi}^\frac{7}{2}}\sinh \left(\frac{4}{|\vec{\beta}_\Earth|}\sqrt{\frac{2T_X}{\Delta E_{fi}}}\right)e^{-\frac{2 T_{X}}{|\vec{\beta}_\Earth|^2 \Delta E_{fi}}}\\
    &= 9.74 (g_L - g_R)^2\, s^{-1},
    \end{split}
\end{equation}
where in going from the first to the second line we made the approximation $T_X \simeq \frac{1}{2}\Delta E_{fi} |\vec{\beta}_\Earth|^2$. This is significantly smaller than the electromagnetic transition rate~\eqref{eq:photonRate} computed in Section~\ref{sec:photonRate}, but should have close to zero background from the forbidden electromagnetic transition $1S_{1/2} \to 2S_{1/2}$, which must proceed through the heavily suppressed magnetic and electric dipole operators. Of course, in order for this transition to occur at all we require that $m_X \simeq \Delta E_{fi}$, however there is no reason \textit{a priori} why we should choose the $n = 1$ to $n' = 2$ transition. Indeed, choosing a transition with smaller gap, would allow us to probe lighter dark matter, with a rate scaling strongly as $\Delta E_{fi}^{-3}$. On the other hand, we know that since $C_1 \sim Z^4$ and $\Delta E_{fi} \sim Z^2$, this transition rate will scale as $Z^{-2}$ overall, suggesting that hydrogen or some other light element is the best choice of target for this process. With that said, targets with a larger $Z$ may still be better suited for heavier dark matter searches, as they typically feature larger $\Delta E_{fi}$. Importantly, \texttt{CINCO} provides the ability to quickly and precisely compute such amplitudes, opening up many possibilities for new physics searches in atomic transitions.

\section{Conclusions}\label{sec:conclusions}

We have presented the first, comprehensive computation of transition amplitudes in hydrogen and hydrogen-like atoms for \emph{generalised hydrogen interactions}. This includes results for a complete set of Lorentz structures corresponding to scalar, pseudoscalar, vector, axial-vector and tensor interactions as well as interference terms in the case in which multiple interactions are present. 

To this end, we introduced a particularly well-suited basis for solving the Dirac equation with a central Coulomb potential, evaluated the angular and radial integrals, and expressed the different transition rates in terms of these integrals for the different Lorentz structures. Further, we have developed a dedicated tool for the construction of hydrogenic transition amplitudes and numerical evaluation of the overlap integrals; \texttt{CINCO}, publicly available at \href{https://gitlab.com/JShergold/cinco}{gitlab.com/JShergold/cinco}. We have demonstrated the precision of our computation by reproducing the QED prediction for the $2P_{3/2} \to 1S_{1/2}$ hydrogen transition to a precision of $\sim 0.005\%$. We have also given an example for a new physics transition by calculating the $1S_{1/2} \to 2S_{1/2}$ absorption rate for a spin-1 dark matter field with both vector and axial-vector couplings to electrons.

Our work is directly relevant to searches for light Dark Matter particles or other light new physics such as axions, hidden photons or scalar fields that can be absorbed or emitted by hydrogen. It is additionally useful for calculating different transition rates in laboratory searches for fifth forces, or variations of fundamental constants. Another important application is the calculation of transition rates in stars or the intergalactic medium, e.g. the Lyman-$\alpha$ forest.

In our work, we have emphasised the power law scaling of atomic transitions with the atomic number that can be as large as $Z^{14}$ for dimension six operators induced by SM processes, or more generally $Z^{2(d-1)}$ for dimension $d$ operators. This incentivises the use of highly charged ion clocks for spectroscopy measurements that would be particularly sensitive to transitions induced by non-renormalisable operators, e.g. the emission or absorption of neutrino-antineutrino pairs mediated by the weak force, or transitions induced by axions with non-renormalisable couplings to electrons.  We will discuss the sensitivity of this proposal to new physics in a dedicated publication.

While this paper comprehensively covers all atomic transitions in hydrogen or hydrogenic ions in the dipole approximation, there are several ways in which our work can be extended. A very interesting direction is to extend our analysis to include generalised hyperfine structure transitions. An orthogonal direction is the analysis of generalised transitions in helium and other multi-electron atoms. While the precision achieved by the relativistic treatment is remarkable, it could be further improved by implementing corrections from the higher multiple terms. Finally, a similar approach can be used to calculate rates for nuclear transitions.
%
%
\acknowledgments
Jack D. Shergold would like to thank Jonas Spinner for his mentorship on good programming practices, which greatly aided in the development of \texttt{CINCO}. Martin Bauer is supported by the UKRI Future Leadership Fellowship DARKMAP. Javier Perez-Soler is supported by the grant CIACIF/2022/158, funded by Generalitat Valenciana. This work is further supported by the Spanish grants PID2020-113775GB-I00 (AEI/10.13039/501100011033) and CIPROM/2021/054 (Generalitat Valenciana).
\appendix
\section{Hydrogen wavefunction and radial integral scaling}\label{sec:wavefunction}
In this appendix, we will present the complete forms of the spinor spherical harmonics and radial wavefunctions obtained by solving the Dirac equation for an electron in a Coulomb potential~\eqref{eq:diracEq}. Additionally, we will briefly derive the leading order scaling of the radial integrals~\eqref{eq:radialIntegrals} with respect to $Z$. 

The spinor spherical harmonics are given by the Clebsch-Gordan decomposition
\begin{equation}
    \Omega_{j,l,m}(\theta,\phi) = \sum_{m_s} C_{m-m_s,m_s,m}^{l,s,j} Y_{l,m-m_s}(\theta,\phi) \xi_{m_s},
\end{equation}
where $s = \frac{1}{2}$ is the spin quantum number, $m_s \in \{-\frac{1}{2}, \frac{1}{2}\}$ is the magnetic spin quantum number, $C$ denotes a Clebsch-Gordan coefficient, $Y_{l,m}$ denotes a spherical harmonic and
\begin{equation}
    \xi_{m_s} = \left(\begin{array}{c}
         \delta_{m_s,\frac{1}{2}}  \\
         \delta_{m_s,-\frac{1}{2}} 
    \end{array}\right)
\end{equation}
is a basis spinor. The radial functions are significantly more complicated, and their full derivation can be found in~\cite{landauQED}. Here we simply quote the results
\begin{equation}
    \begin{split}
    f_{n,\kappa}(r),g_{n,\kappa}(r) &= \pm \frac{(2\lambda_{n,\kappa})^\frac{3}{2}}{\Gamma(2\gamma_\kappa + 1)}\left[\frac{(\mu \pm E_{n,\kappa})\Gamma(2\gamma_{\kappa} + N_{n,\kappa} + 1)}{4\mu\left(\frac{Z\alpha_\mathrm{EM} \mu}{\lambda_{n,\kappa}}\right)\left(\frac{Z\alpha_\mathrm{EM} \mu}{\lambda_{n,\kappa}} - \kappa\right)N_{n,\kappa}!}\right]^\frac{1}{2}\rho_{n,\kappa}^{\gamma_\kappa -1} e^{-\frac{\rho_{n,\kappa}}{2}} \\
    &\times\Bigg[\left(\frac{Z\alpha_\mathrm{EM} \mu}{\lambda_{n,\kappa}} - \kappa\right){_1F_1}(-N_{n,\kappa},2\gamma_\kappa+1,\rho_{n,\kappa}) \\
    &\qquad\qquad\qquad\qquad\qquad\qquad\mp N_{n,\kappa} {_1F_1}(1-N_{n,\kappa},2\gamma_\kappa+1,\rho_{n,\kappa})\Bigg],
    \end{split}
\end{equation}
where 
\begin{equation}
    \begin{alignedat}{3}
    \gamma_\kappa &= \sqrt{\kappa^2 - (Z\alpha_\mathrm{EM})^2}, \qquad &\lambda_{n,\kappa} &= \sqrt{\mu^2 - E_{n,\kappa}^2},\\
    N_{n,\kappa} &= n - |\kappa| \in \{0,1,2,\dots\},\qquad &\rho_{n,\kappa} &= 2\lambda_{n,\kappa} r,
    \end{alignedat}
\end{equation}
and the energy of the state is given by
\begin{equation}
    E_{n,\kappa} = \frac{\mu}{\sqrt{1 + \left(\frac{Z\alpha_\mathrm{EM}}{\gamma_\kappa + N_{n,\kappa}}\right)^2}} \simeq \mu \left[1 - \frac{1}{2}\left(\frac{Z\alpha_\mathrm{EM}}{n}\right)^2 + \mathcal{O}\left(\alpha_\mathrm{EM}^4\right)\right].
\end{equation}
Here, the second term in the expansion is proportional to the Rydberg constant, $R_\infty = \alpha_\mathrm{EM}^2 \mu/2$, recovering the non-relativistic result. The confluent hypergeometric functions can be expressed in terms of generalised Laguerre polynomials
\begin{equation}
    {_1F_1}(-m,a,z) = \frac{m!}{(a)_m} L_{m}^{(a-1)}(z),
\end{equation}
for $m \geq 0$, where $(a)_m$ denotes the Pochhammer symbol. However, a more useful expression for our purposes is the terminating power series 
\begin{equation}\label{eq:1F1series}
    {_1F_1}(-m,a,z) = \sum_{n=0}^{m} (-1)^n \binom{m}{n}\frac{z^n}{(a)_n},
\end{equation}
where $\binom{m}{n}$ is the binomial coefficient. Combining~\eqref{eq:1F1series} with the identity
\begin{equation}
    \int_0^\infty dr\,r^n e^{-ar} = \frac{1}{a^{n+1}} \Gamma(n+1),
\end{equation}
allows us to find the leading order scaling of the radial integrals~\eqref{eq:radialIntegrals} with respect to $Z$
\begin{equation}
    \mathcal{I}_{ff} \sim Z^{0}, \qquad \mathcal{I}_{fg} \sim \mathcal{I}_{gf} \sim Z^{1}, \qquad \mathcal{I}_{gg} \sim Z^{2}.
\end{equation}
These results follow from the product of the integral
\begin{equation}
    \begin{split}
    \int_0^\infty dr\, r^2 \rho_{n,\kappa}^{\gamma_\kappa - 1} \rho_{n',\kappa'}^{\gamma_\kappa' - 1} &e^{-\frac{1}{2}(\rho_{n,\kappa} +\rho_{n',\kappa'})}{_1F_1}(m,2\gamma_{\kappa}+1,\rho_{n,\kappa}){_1F_1}(m',2\gamma_{\kappa'}+1,\rho_{n',\kappa'}) \sim Z^{-3},
    \end{split}
\end{equation}
with the $r$-independent prefactors in $f$ and $g$, which to leading order scale as $Z^\frac{3}{2}$ and $Z^{\frac{5}{2}}$, respectively.

%
%
\section{Dipole approximation}\label{sec:dipoleApproximation}
When computing the transition amplitude~\eqref{eq:ampSq}, we noted that in most cases the phase factor can be replaced with one, corresponding to the dipole approximation. In this appendix we will justify this approximation. The typical scale of an atom, or stated differently, the radius at which you are most likely to find a hydrogen electron is
\begin{equation}
    \left\langle r_n \right\rangle \simeq \frac{a_0 n^2}{Z} = \frac{n^2}{Z} (2.68\cdot 10^{-4}\,\mathrm{eV}^{-1}),
\end{equation}
where $n$ is the principal quantum number and $a_0$ is the Bohr radius. Additionally, the momenta of the absorbed or emitted particles will be approximately
\begin{equation}
    \left\langle k \right\rangle \simeq \beta E,
\end{equation}
where $\beta$ is their velocity and $E$ is their energy, which should be of order the transition energy that we are interested in
\begin{equation}
    E\simeq \Delta E_{fi} \simeq R_\infty Z^2 \left|\frac{1}{n^2} - \frac{1}{n'^2} \right| = Z^2 \left|\frac{1}{n^2} - \frac{1}{n'^2} \right| \left(13.6\,\mathrm{eV}\right),
\end{equation}
with $R_\infty$ the Rydberg constant and $n'$ the principal quantum number of the final state. The requirement that $\vec{k} \cdot \vec{x} \ll 1$ therefore translates to
\begin{equation}\label{eq:validity}
    \beta Z\left|1-\left(\frac{n}{n'}\right)^2\right| \ll 274.
\end{equation}
In the case of absorption, $n' \geq n$ and the LHS is maximised for $n' \gg n$, in which case we require that
\begin{equation}
    \beta Z \ll 274, \quad \mathrm{absorption},
\end{equation}
which should hold even for relativistic particles with $\beta = 1$. On the other hand, if we are considering emission with $n \geq n'$, then the LHS is maximised for $n \gg n'$ and we instead require that
\begin{equation}
    \beta Z \left(\frac{n}{n'}\right)^2 \ll 274, \quad \mathrm{emission}.
\end{equation}
For non-relativistic particles \textit{e.g.} dark matter with $\beta = 10^{-3}$, and reference $Z \simeq 10$, this holds for $n \ll 150 n'$, which should be satisfied in all but the most extreme cases. For relativistic particles, this becomes much stricter and we require that $n \ll 5n'$, which still covers most realistic cases.  We stress, however, that the validity should be checked with~\eqref{eq:validity} before using these amplitudes.
%
%
\section{Spherical basis}\label{sec:spherical-basis}
In this appendix we will briefly introduce the spherical basis, and give useful relations for computing hydrogen amplitudes. We begin by defining the components of a vector in the spherical basis in terms of their Cartesian components as
\begin{equation}
    V^{\pm} = \mp \frac{1}{\sqrt{2}}(V^x \pm i V^y),
\end{equation}
so as to align with the definitions of the raising and lowering operators. To define the inner product and covariant four-vector components in the spherical basis we will also need the metric tensor. This can be found by making use of the relation
\begin{equation}
    g_{\mu\nu} = (\Lambda^{-1})^{\rho}_{\ \mu}(\Lambda^{-1})^{\sigma}_{\ \nu} g'_{\rho\sigma},
\end{equation}
where for the remainder of this appendix, primed variables will denote a quantity in Cartesian coordinates, and where
\begin{equation}
     \left(\Lambda^{-1}\right)^{\rho}_{\ \mu} = \pderiv{x^{\prime\rho}}{x^\mu} = \begin{pNiceMatrix}[columns-width = auto]
         1 & 0 & 0 & 0 \\
         0 & -\frac{1}{\sqrt{2}} & \frac{1}{\sqrt{2}} & 0 \\
         0 & \frac{i}{\sqrt{2}} & \frac{i}{\sqrt{2}} & 0 \\
         0 & 0 & 0 & 1
     \end{pNiceMatrix},
\end{equation}
are the components of the inverse coordinate transform matrix, with row indices $(0, x, y, z)$ from top to bottom, and column indices $(0, +, -, z)$ from left to right. Using this, we find the components of the metric in the spherical basis
\begin{equation}
    g_{\mu\nu} = \begin{pNiceMatrix}[columns-width=auto]
        1 & 0 & 0 & 0  \\
        0 & 0 & 1 & 0 \\
        0 & 1 & 0 & 0 \\
        0 & 0 & 0 & -1
    \end{pNiceMatrix},
\end{equation}
with both row and column indices in the order $(0, +, -, z)$. Using the metric, we can define the covariant components of a four-vector
\begin{equation}
    V_\mu = g_{\mu\nu} V^\nu \implies V_0 = V^0, \qquad V_+ = V^-, \qquad V_- = V^+, \qquad V_z = -V^z.
\end{equation}
As an example, we show what the on-shell condition looks like in the spherical basis. The contravariant components of the four-momentum in the spherical basis are
\begin{equation}\label{eq:onshell}
    p^\mu = \left(\begin{array}{c}
         p^0  \\
         p^+ \\
         p^-\\
         p^z
    \end{array}\right) = \left(\begin{array}{c}
         E  \\
        -\frac{p}{\sqrt{2}}\sin\theta e^{i\phi}\\
        \frac{p}{\sqrt{2}}\sin\theta e^{-i\phi} \\
         p \cos\theta
    \end{array}\right),
\end{equation}
where $\theta$ and $\phi$ are the polar and azimuthal angles, and $p$ is the modulus of its three-momentum. It then follows that
\begin{equation}
    \begin{split}
    p_\mu p^\mu &= p_0 p^0 + p_+ p^+ + p_- p^- + p_z p^z \\
    &= E^2 - p^2\sin^2\theta - p^2\cos\theta\\
    & = m^2,
    \end{split}
\end{equation}
as we expect. In general, a quantity in the spherical basis can be found from the one in the Cartesian basis by making use of
\begin{equation}
    F_{\mu\nu\dots}^{\alpha\beta\dots} = \Lambda_{\mu}^{\ \sigma}\Lambda_{\nu}^{\ \rho}\dots  (\Lambda^{-1})^{\alpha}_{\ \kappa}(\Lambda^{-1})^{\beta}_{\ \psi} \dots F^{\prime\kappa\psi\dots}_{\rho\sigma\dots},
\end{equation}
where
\begin{equation}
    \Lambda^{\ \alpha}_{\kappa} = \pderiv{x^\alpha}{x^{\prime\kappa}} = \begin{pNiceMatrix}[columns-width = auto]
         1 & 0 & 0 & 0 \\
         0 & -\frac{1}{\sqrt{2}} & -\frac{i}{\sqrt{2}} & 0 \\
         0 & \frac{1}{\sqrt{2}} & -\frac{i}{\sqrt{2}} & 0 \\
         0 & 0 & 0 & 1
     \end{pNiceMatrix}\;, \qquad \Lambda^{\ \alpha}_{\kappa} (\Lambda^{-1})_{\ \beta}^{\kappa} = \delta^{\alpha}_\beta\;,
\end{equation}
are the components of the transformation matrix from the spherical to Cartesian basis, again with row indices $(0,x,y,z)$ and column indices $(0,+,-,z)$. As such, only fully contracted objects such as~\eqref{eq:onshell} have the same functional form in both bases. Of particular interest to us are the gamma matrices in the spherical Dirac basis, which are
\begin{equation}
    \gamma^{\pm} = \begin{pNiceMatrix}
        0 & \sigma^{\pm} \\
        -\sigma^{\pm} & 0
    \end{pNiceMatrix}, \qquad \sigma^\pm = \mp\frac{1}{\sqrt{2}}(\sigma^x \pm i\sigma^y)
\end{equation}
and the tensor operator
\begin{equation}
    \sigma^{\mu\nu} = \Lambda^{\ \mu}_{\alpha}\Lambda^{\ \nu}_{\beta} \left(\frac{i}{2}[\gamma^{\prime\alpha},\gamma^{\prime\beta}]\right) =\frac{i}{2}[\gamma^{\mu},\gamma^{\nu}]. 
\end{equation}
\section{Transition tables}\label{sec:transitionTables}
In this appendix we tabulate the full set of allowed transitions for each Lorentz structure and index in the spherical basis, along with their selection rules, radial integral structure and coefficient. Each entry of a table corresponds to one of the integrals 
\begin{equation}
    \int d^3x\, \bar{\mathcal{U}}_{n',\kappa',m'} \Gamma^{\{\mu\}}\mathcal{U}_{n,\kappa,m} = \left(\sum_t C_{\kappa,m}^{(t)} \delta_{t} \mathcal{I}_t\right),
\end{equation}
which form part of the atomic tensor~\eqref{eq:atomicTensor}. In all three tables, $\Gamma$ denotes the Lorentz structure with Lorentz indices $\{\mu\}$, $\Delta m$ and $\Delta \kappa$ denote the change in the magnetic quantum number, $m$, and $\kappa$, respectively, $\mathcal{I}$ denotes the radial integral structure, and $C_{\kappa,m}$ denotes the coefficient of radial integral that arises from the angular integrals.
\vspace*{\fill}
\begin{table}[h!]
\setcellgapes{5pt}
\makegapedcells
\renewcommand\multirowsetup{\centering}
\setlength\tabcolsep{3mm}
\centering
\begin{tabular}{Sc|Sc|Sc|Sc|Sc|Sc}
$\Gamma^{\{\mu\}}$ & ${\{\mu\}}$ & $\Delta m$ & $\Delta \kappa$ & $\mathcal{I}$ & $C_{\kappa,m}$ \\ \hline\hline
$1$ & - & $m \to m$ & $\kappa \to \kappa$ & $\mathcal{I}_{ff} - \mathcal{I}_{gg}$ & $1$ \\ \hline
$\gamma^5$ & - & $m \to m$ & $\kappa \to -\kappa$ & $\mathcal{I}_{fg} + \mathcal{I}_{gf}$ & $i$
\end{tabular}
\caption{Allowed transitions, radial integrals, and their coefficients for the scalar and pseudoscalar vertices. }
\label{tab:transitions-scalar}
\end{table}
\vspace*{\fill}

\clearpage
\begin{table}[h!]
\setcellgapes{3.5pt}
\makegapedcells
\renewcommand\multirowsetup{\centering}
\setlength\tabcolsep{3mm}
\hspace{-1.2cm}
\begin{tabular}{Sc|Sc|Sc|Sc|Sc|Sc}
$\Gamma^{\{\mu\}}$ & ${\{\mu\}}$ & $\Delta m$ & $\Delta \kappa$ & $\mathcal{I}$ & $C_{\kappa,m}$ \\ \hline\hline
\multirow{19.5}{*}{$\gamma^\mu$} & $0$ & $m \to m$ & $\kappa \to \kappa$ & $\mathcal{I}_{ff} + \mathcal{I}_{gg} (=0)$ & $1$ \\ \cline{2-6} 
 & \multirow{8.1}{*}{$\pm$} & \multirow{8.1}{*}{$m \to m \pm 1$} & $\kappa \to - \kappa$ & $\dfrac{\mathcal{I}_{fg}}{2\kappa-1} + \dfrac{\mathcal{I}_{gf}}{2\kappa+1}$ & $\mp i \sqrt{2} \sqrt{\kappa^2 - \left(m\pm \frac{1}{2}\right)^2}$ \\ \cline{4-6} 
 &  &  & $\kappa \to \kappa + 1$ & $\mathcal{I}_{gf}$ & $\dfrac{i \sqrt{2} \sqrt{\left(\kappa\pm m+\frac{1}{2}\right)\left(\kappa \pm m+\frac{3}{2}\right)}}{2 \kappa+1}$ \\ \cline{4-6} 
 &  &  & $\kappa \to \kappa - 1$ & $\mathcal{I}_{fg}$ & $\dfrac{i \sqrt{2} \sqrt{\left(\kappa \mp m-\frac{1}{2}\right)\left(\kappa\mp m - \frac{3}{2}\right)}}{2 \kappa-1}$ \\ \cline{2-6} 
 & \multirow{8}{*}{$z$} & \multirow{8}{*}{$m \to m$} & $\kappa \to - \kappa$ & $\dfrac{\mathcal{I}_{fg}}{2\kappa-1} + \dfrac{\mathcal{I}_{gf}}{2\kappa+1}$ & $2im$ \\ \cline{4-6} 
 &  &  & $\kappa \to \kappa + 1$ & $\mathcal{I}_{gf}$ & $\dfrac{2i\sqrt{\left(\kappa + \frac{1}{2}\right)^2 - m^2}}{|2\kappa+1|}$ \\ \cline{4-6} 
 &  &  & $\kappa \to \kappa - 1$ & $\mathcal{I}_{fg}$ & $-\dfrac{2i\sqrt{\left(\kappa - \frac{1}{2}\right)^2 - m^2}}{|2\kappa-1|}$ \\ \hline
\multirow{19.5}{*}{$\gamma^\mu \gamma^5$} & $0$ & $m \to m$ & $\kappa \to - \kappa$ & $\mathcal{I}_{fg} - \mathcal{I}_{gf}$ & $i$ \\ \cline{2-6} 
 & \multirow{8.1}{*}{$\pm$} & \multirow{8.1}{*}{$m \to m \pm 1$} & $\kappa \to \kappa$ & $\dfrac{\mathcal{I}_{ff}}{2\kappa+1} - \dfrac{\mathcal{I}_{gg}}{2\kappa-1}$ & $\pm \sqrt{2} \sqrt{\kappa^2 - \left(m\pm \frac{1}{2}\right)^2}$ \\ \cline{4-6} 
 &  &  & $\kappa \to -\kappa + 1$ & $\mathcal{I}_{gg}$ & $\dfrac{\sqrt{2} \sqrt{\left(\kappa \mp m-\frac{1}{2}\right)\left(\kappa\mp m - \frac{3}{2}\right)}}{2 \kappa-1}$ \\ \cline{4-6} 
 &  &  & $\kappa \to -\kappa - 1$ & $\mathcal{I}_{ff}$ & $-\dfrac{\sqrt{2} \sqrt{\left(\kappa \pm m+\frac{1}{2}\right)\left(\kappa\pm m + \frac{3}{2}\right)}}{2 \kappa+1}$ \\ \cline{2-6} 
 & \multirow{8}{*}{$z$} & \multirow{8}{*}{$m \to m$} & $\kappa \to \kappa$ & $\dfrac{\mathcal{I}_{ff}}{2\kappa+1} - \dfrac{\mathcal{I}_{gg}}{2\kappa-1}$ & $-2m$ \\ \cline{4-6} 
 &  &  & $\kappa \to -\kappa + 1$ & $\mathcal{I}_{gg}$ & $-\dfrac{2\sqrt{\left(\kappa - \frac{1}{2}\right)^2 - m^2}}{|2\kappa-1|}$ \\ \cline{4-6} 
 &  &  & $\kappa \to -\kappa - 1$ & $\mathcal{I}_{ff}$ & $-\dfrac{2\sqrt{\left(\kappa + \frac{1}{2}\right)^2 - m^2}}{|2\kappa+1|}$
\end{tabular}
\caption{Allowed transitions, radial integrals, and their coefficients for the vector and axial-vector vertices. Note that the $\gamma^0$ vertex is identically zero due to wavefunction orthogonality.}
\label{tab:transitions-vector}
\end{table}
\clearpage
\begin{table}[h!]
\setcellgapes{5pt}
\makegapedcells
\renewcommand\multirowsetup{\centering}
\setlength\tabcolsep{3mm}
\hspace{-1.2cm}
\begin{tabular}{Sc|Sc|Sc|Sc|Sc|Sc}
$\Gamma^{\{\mu\}}$ & ${\{\mu\}}$ & $\Delta m$ & $\Delta \kappa$ & $\mathcal{I}$ & $C_{\kappa,m}$ \\ \hline\hline
\multirow{40.5}{*}{$\sigma^{\mu\nu}$} & \multirow{7}{*}{$0\pm$} & \multirow{7}{*}{$m \to m\pm 1$} & $\kappa \to -\kappa$ & $\dfrac{\mathcal{I}_{fg}}{2\kappa-1} - \dfrac{\mathcal{I}_{gf}}{2\kappa+1}$ & $\pm\sqrt{2}\sqrt{\kappa^2 - \left(m\pm \frac{1}{2}\right)^2}$ \\ \cline{4-6} 
 &  &  & $\kappa \to \kappa +1$ & $\mathcal{I}_{gf}$ & $\dfrac{\sqrt{2}\sqrt{\left(\kappa\pm m + \frac{1}{2}\right)\left(\kappa \pm m + \frac{3}{2}\right)}}{2\kappa+1}$ \\ \cline{4-6} 
 &  &  & $\kappa \to \kappa - 1$ & $\mathcal{I}_{fg}$ & $-\dfrac{\sqrt{2}\sqrt{\left(\kappa\mp m - \frac{1}{2}\right)\left(\kappa \mp m - \frac{3}{2}\right)}}{2\kappa-1}$ \\ \cline{2-6} 
 & \multirow{7}{*}{$0z$} & \multirow{7}{*}{$m\to m$} & $\kappa \to -\kappa$ & $\dfrac{\mathcal{I}_{fg}}{2\kappa -1} - \dfrac{\mathcal{I}_{gf}}{2\kappa +1}$ & $-2m$ \\ \cline{4-6} 
 &  &  & $\kappa \to \kappa + 1$ & $\mathcal{I}_{gf}$ & $\dfrac{2\sqrt{\left(\kappa+\frac{1}{2}\right)^2 - m^2}}{|2\kappa+1|}$ \\ \cline{4-6} 
 &  &  & $\kappa \to \kappa -1$ & $\mathcal{I}_{fg}$ & $\dfrac{2\sqrt{\left(\kappa-\frac{1}{2}\right)^2 - m^2}}{|2\kappa-1|}$ \\ \cline{2-6} 
 & \multirow{7}{*}{$+-$} & \multirow{7}{*}{$m\to m$} & $\kappa \to \kappa$ & $\dfrac{\mathcal{I}_{ff}}{2\kappa+1} + \dfrac{\mathcal{I}_{gg}}{2\kappa-1}$ & $-2im$ \\ \cline{4-6} 
 &  &  & $\kappa \to -\kappa +1$ & $\mathcal{I}_{gg}$ & $\dfrac{2i\sqrt{\left(\kappa - \frac{1}{2}\right)^2 - m^2}}{|2\kappa-1|}$ \\ \cline{4-6} 
 &  &  & $\kappa \to -\kappa -1$ & $\mathcal{I}_{ff}$ & $-\dfrac{2i\sqrt{\left(\kappa + \frac{1}{2}\right)^2 - m^2}}{|2\kappa+1|}$ \\ \cline{2-6} 
 & \multirow{7.2}{*}{$\pm z$} & \multirow{7.2}{*}{$m\to m\pm1$} & $\kappa \to \kappa$ & $\dfrac{\mathcal{I}_{ff}}{2\kappa +1} + \dfrac{\mathcal{I}_{gg}}{2\kappa-1}$ & $-i\sqrt{2}\sqrt{\kappa^2 - \left(m \pm \frac{1}{2}\right)^2}$ \\ \cline{4-6} 
 &  &  & $\kappa \to -\kappa +1$ & $\mathcal{I}_{gg}$ & $\mp\dfrac{i\sqrt{2}\sqrt{\left(\kappa \mp m - \frac{1}{2}\right)\left(\kappa \mp m -\frac{3}{2}\right)}}{2\kappa-1}$ \\ \cline{4-6} 
 &  &  & $\kappa \to -\kappa -1$ & $\mathcal{I}_{ff}$ & $\mp\dfrac{i\sqrt{2}\sqrt{\left(\kappa \pm m + \frac{1}{2}\right)\left(\kappa \pm m +\frac{3}{2}\right)}}{2\kappa+1}$
\end{tabular}
\caption{Allowed transitions, radial integrals, and their coefficients for the tensor vertices, where we have not included the zero diagonal vertices, and where any vertices with permuted indices can be found from the antisymmetry of the tensor operator $\sigma^{\mu\nu}$.}
\label{tab:transitions-tensor}
\end{table}
\clearpage
%
%
\bibliographystyle{JHEP}
\bibliography{bibliography}

\providecommand{\href}[2]{#2}\begingroup\raggedright\begin{thebibliography}{10}

\bibitem{RevModPhys.28.53}
H.~E. Suess and H.~C. Urey, \emph{Abundances of the elements}, \href{https://doi.org/10.1103/RevModPhys.28.53}{\emph{Rev. Mod. Phys.} {\bfseries 28} (Jan, 1956) 53--74}.

\bibitem{ANDERS19822363}
E.~Anders and M.~Ebihara, \emph{Solar-system abundances of the elements}, \href{https://doi.org/https://doi.org/10.1016/0016-7037(82)90208-3}{\emph{Geochimica et Cosmochimica Acta} {\bfseries 46} (1982) 2363--2380}.

\bibitem{Lodders:2003vvq}
K.~Lodders, \emph{{Solar System Abundances and Condensation Temperatures of the Elements}}, \href{https://doi.org/10.1086/375492}{\emph{Astrophys. J.} {\bfseries 591} (2003) 1220--1247}.

\bibitem{PhysRev.151.1189}
G.~Steigman, \emph{Neutrino pair production in bound-bound transitions}, \href{https://doi.org/10.1103/PhysRev.151.1189}{\emph{Phys. Rev.} {\bfseries 151} (Nov, 1966) 1189--1191}.

\bibitem{Huang:2019rmc}
G.-Y. Huang and S.~Zhou, \emph{{Probing Cosmic Axions through Resonant Emission and Absorption in Atomic Systems with Superradiance}}, \href{https://doi.org/10.1103/PhysRevD.100.035010}{\emph{Phys. Rev. D} {\bfseries 100} (2019) 035010}, [\href{https://arxiv.org/abs/1905.00367}{{\ttfamily 1905.00367}}].

\bibitem{Donnelly:1978ty}
T.~W. Donnelly, S.~J. Freedman, R.~S. Lytel, R.~D. Peccei and M.~Schwartz, \emph{{Do Axions Exist?}}, \href{https://doi.org/10.1103/PhysRevD.18.1607}{\emph{Phys. Rev. D} {\bfseries 18} (1978) 1607}.

\bibitem{Sikivie:2014lha}
P.~Sikivie, \emph{{Axion Dark Matter Detection using Atomic Transitions}}, \href{https://doi.org/10.1103/PhysRevLett.113.201301}{\emph{Phys. Rev. Lett.} {\bfseries 113} (2014) 201301}, [\href{https://arxiv.org/abs/1409.2806}{{\ttfamily 1409.2806}}].

\bibitem{Santamaria:2015gro}
L.~Santamaria, C.~Braggio, G.~Carugno, V.~D. Sarno, P.~Maddaloni and G.~Ruoso, \emph{{Axion dark matter detection by laser spectroscopy of ultracold molecular oxygen: a proposal}}, \href{https://doi.org/10.1088/1367-2630/17/11/113025}{\emph{New J. Phys.} {\bfseries 17} (2015) 113025}.

\bibitem{Yang:2019xdz}
Q.~Yang and S.~Dong, \emph{{Probing dark matter axions using the hyperfine structure splitting of hydrogen atoms}}, \href{https://doi.org/10.1016/j.physletb.2023.138004}{\emph{Phys. Lett. B} {\bfseries 843} (2023) 138004}, [\href{https://arxiv.org/abs/1912.11472}{{\ttfamily 1912.11472}}].

\bibitem{Vergados:2024vky}
J.~D. Vergados, S.~Cohen, F.~T. Avignone and R.~Creswick, \emph{{Searching for Dark Matter Axions via Atomic Excitations}}, \href{https://doi.org/10.3390/particles7010006}{\emph{Particles} {\bfseries 7} (2024) 96--120}.

\bibitem{Flambaum:2018wbu}
V.~V. Flambaum, I.~B. Samsonov, H.~B. Tran~Tan and D.~Budker, \emph{{Coherent axion-photon transformations in the forward scattering on atoms}}, \href{https://doi.org/10.1103/PhysRevD.98.095028}{\emph{Phys. Rev. D} {\bfseries 98} (2018) 095028}, [\href{https://arxiv.org/abs/1805.01793}{{\ttfamily 1805.01793}}].

\bibitem{TranTan:2018bch}
H.~B. Tran~Tan, V.~V. Flambaum, I.~B. Samsonov, Y.~V. Stadnik and D.~Budker, \emph{{Interference-assisted resonant detection of axions}}, \href{https://doi.org/10.1016/j.dark.2019.100272}{\emph{Phys. Dark Univ.} {\bfseries 24} (2019) 100272}, [\href{https://arxiv.org/abs/1803.09388}{{\ttfamily 1803.09388}}].

\bibitem{Yang:2016zaz}
Q.~Yang and H.~Di, \emph{{Vector Dark Matter Detection using the Quantum Jump of Atoms}}, \href{https://doi.org/10.1016/j.physletb.2018.03.045}{\emph{Phys. Lett. B} {\bfseries 780} (2018) 622--626}, [\href{https://arxiv.org/abs/1606.01492}{{\ttfamily 1606.01492}}].

\bibitem{Alvarez-Luna:2018jsb}
C.~\'Alvarez-Luna and J.~A.~R. Cembranos, \emph{{Dark photon searches with atomic transitions}}, \href{https://doi.org/10.1007/JHEP07(2019)110}{\emph{JHEP} {\bfseries 07} (2019) 110}, [\href{https://arxiv.org/abs/1812.08501}{{\ttfamily 1812.08501}}].

\bibitem{Arvanitaki:2017nhi}
A.~Arvanitaki, S.~Dimopoulos and K.~Van~Tilburg, \emph{{Resonant absorption of bosonic dark matter in molecules}}, \href{https://doi.org/10.1103/PhysRevX.8.041001}{\emph{Phys. Rev. X} {\bfseries 8} (2018) 041001}, [\href{https://arxiv.org/abs/1709.05354}{{\ttfamily 1709.05354}}].

\bibitem{Auriol:2018ovo}
A.~Auriol, S.~Davidson and G.~Raffelt, \emph{{Axion absorption and the spin temperature of primordial hydrogen}}, \href{https://doi.org/10.1103/PhysRevD.99.023013}{\emph{Phys. Rev. D} {\bfseries 99} (2019) 023013}, [\href{https://arxiv.org/abs/1808.09456}{{\ttfamily 1808.09456}}].

\bibitem{Fortier:2007jf}
T.~M. Fortier et~al., \emph{{Precision atomic spectroscopy for improved limits on variation of the fine structure constant and local position invariance}}, \href{https://doi.org/10.1103/PhysRevLett.98.070801}{\emph{Phys. Rev. Lett.} {\bfseries 98} (2007) 070801}.

\bibitem{Rosenband:2008qgq}
T.~Rosenband et~al., \emph{{Frequency Ratio of Al+ and Hg+ Single-Ion Optical Clocks; Metrology at the 17th Decimal Place}}, \href{https://doi.org/10.1126/science.1154622}{\emph{Science} {\bfseries 319} (2008) 1154622}.

\bibitem{Derevianko:2013oaa}
A.~Derevianko and M.~Pospelov, \emph{{Hunting for topological dark matter with atomic clocks}}, \href{https://doi.org/10.1038/nphys3137}{\emph{Nature Phys.} {\bfseries 10} (2014) 933}, [\href{https://arxiv.org/abs/1311.1244}{{\ttfamily 1311.1244}}].

\bibitem{Arvanitaki:2014faa}
A.~Arvanitaki, J.~Huang and K.~Van~Tilburg, \emph{{Searching for dilaton dark matter with atomic clocks}}, \href{https://doi.org/10.1103/PhysRevD.91.015015}{\emph{Phys. Rev. D} {\bfseries 91} (2015) 015015}, [\href{https://arxiv.org/abs/1405.2925}{{\ttfamily 1405.2925}}].

\bibitem{Godun:2014naa}
R.~M. Godun, P.~B.~R. Nisbet-Jones, J.~M. Jones, S.~A. King, L.~A.~M. Johnson, H.~S. Margolis et~al., \emph{{Frequency Ratio of Two Optical Clock Transitions in Yb+171 and Constraints on the Time Variation of Fundamental Constants}}, \href{https://doi.org/10.1103/PhysRevLett.113.210801}{\emph{Phys. Rev. Lett.} {\bfseries 113} (2014) 210801}, [\href{https://arxiv.org/abs/1407.0164}{{\ttfamily 1407.0164}}].

\bibitem{2014PhRvL.113u0802H}
N.~Huntemann, B.~Lipphardt, C.~Tamm, V.~Gerginov, S.~Weyers and E.~Peik, \emph{{Improved limit on a temporal variation of $m_p/m_e$ from comparisons of Yb$^+$ and Cs atomic clocks}}, \href{https://doi.org/10.1103/PhysRevLett.113.210802}{\emph{Phys. Rev. Lett.} {\bfseries 113} (2014) 210802}, [\href{https://arxiv.org/abs/1407.4408}{{\ttfamily 1407.4408}}].

\bibitem{Salumbides:2013dua}
E.~J. Salumbides, W.~Ubachs and V.~I. Korobov, \emph{{Bounds on fifth forces at the sub-Angstrom length scale}}, \href{https://doi.org/10.1016/j.jms.2014.04.003}{\emph{J. Molec. Spectrosc.} {\bfseries 300} (2014) 65}, [\href{https://arxiv.org/abs/1308.1711}{{\ttfamily 1308.1711}}].

\bibitem{King:63}
W.~H. King, \emph{{Isotope Shifts in Atomic Spectra}}.
\newblock Physics of Atoms and Molecules. Springer, New York, 1984.

\bibitem{Delaunay:2017dku}
C.~Delaunay, C.~Frugiuele, E.~Fuchs and Y.~Soreq, \emph{{Probing new spin-independent interactions through precision spectroscopy in atoms with few electrons}}, \href{https://doi.org/10.1103/PhysRevD.96.115002}{\emph{Phys. Rev. D} {\bfseries 96} (2017) 115002}, [\href{https://arxiv.org/abs/1709.02817}{{\ttfamily 1709.02817}}].

\bibitem{Flambaum:2017onb}
V.~V. Flambaum, A.~J. Geddes and A.~V. Viatkina, \emph{{Isotope shift, nonlinearity of King plots, and the search for new particles}}, \href{https://doi.org/10.1103/PhysRevA.97.032510}{\emph{Phys. Rev. A} {\bfseries 97} (2018) 032510}, [\href{https://arxiv.org/abs/1709.00600}{{\ttfamily 1709.00600}}].

\bibitem{Berengut:2017zuo}
J.~C. Berengut et~al., \emph{{Probing New Long-Range Interactions by Isotope Shift Spectroscopy}}, \href{https://doi.org/10.1103/PhysRevLett.120.091801}{\emph{Phys. Rev. Lett.} {\bfseries 120} (2018) 091801}, [\href{https://arxiv.org/abs/1704.05068}{{\ttfamily 1704.05068}}].

\bibitem{Solaro:2020dxz}
C.~Solaro, S.~Meyer, K.~Fisher, J.~C. Berengut, E.~Fuchs and M.~Drewsen, \emph{{Improved isotope-shift-based bounds on bosons beyond the Standard Model through measurements of the $^2$D$_{3/2} - ^2$D$_{5/2}$ interval in Ca$^+$}}, \href{https://doi.org/10.1103/PhysRevLett.127.029901}{\emph{Phys. Rev. Lett.} {\bfseries 125} (2020) 123003}, [\href{https://arxiv.org/abs/2005.00529}{{\ttfamily 2005.00529}}].

\bibitem{Door:2024qqz}
M.~Door et~al., \emph{{Search for new bosons with ytterbium isotope shifts}},  \href{https://arxiv.org/abs/2403.07792}{{\ttfamily 2403.07792}}.

\bibitem{Jones:2019qny}
M.~P.~A. Jones, R.~M. Potvliege and M.~Spannowsky, \emph{{Probing new physics using Rydberg states of atomic hydrogen}}, \href{https://doi.org/10.1103/PhysRevResearch.2.013244}{\emph{Phys. Rev. Res.} {\bfseries 2} (2020) 013244}, [\href{https://arxiv.org/abs/1909.09194}{{\ttfamily 1909.09194}}].

\bibitem{Brzeminski:2022sde}
D.~Brzeminski, Z.~Chacko, A.~Dev, I.~Flood and A.~Hook, \emph{{Searching for a fifth force with atomic and nuclear clocks}}, \href{https://doi.org/10.1103/PhysRevD.106.095031}{\emph{Phys. Rev. D} {\bfseries 106} (2022) 095031}, [\href{https://arxiv.org/abs/2207.14310}{{\ttfamily 2207.14310}}].

\bibitem{Delaunay:2016brc}
C.~Delaunay, R.~Ozeri, G.~Perez and Y.~Soreq, \emph{{Probing Atomic Higgs-like Forces at the Precision Frontier}}, \href{https://doi.org/10.1103/PhysRevD.96.093001}{\emph{Phys. Rev. D} {\bfseries 96} (2017) 093001}, [\href{https://arxiv.org/abs/1601.05087}{{\ttfamily 1601.05087}}].

\bibitem{Bauer:2022rwf}
M.~Bauer, G.~Rostagni and J.~Spinner, \emph{{Axion-Higgs portal}}, \href{https://doi.org/10.1103/PhysRevD.107.015007}{\emph{Phys. Rev. D} {\bfseries 107} (2023) 015007}, [\href{https://arxiv.org/abs/2207.05762}{{\ttfamily 2207.05762}}].

\bibitem{Dzuba:2017cas}
V.~A. Dzuba, V.~V. Flambaum, P.~Munro-Laylim and Y.~V. Stadnik, \emph{{Probing Long-Range Neutrino-Mediated Forces with Atomic and Nuclear Spectroscopy}}, \href{https://doi.org/10.1103/PhysRevLett.120.223202}{\emph{Phys. Rev. Lett.} {\bfseries 120} (2018) 223202}, [\href{https://arxiv.org/abs/1711.03700}{{\ttfamily 1711.03700}}].

\bibitem{Zhou:2014xbw}
W.-P. Zhou, P.~Zhou and H.-X. Qiao, \emph{{Detecting extra dimensions by Hydrogen-like atoms}}, \href{https://doi.org/10.1515/phys-2015-0011}{\emph{Open Phys.} {\bfseries 13} (2015) 96--99}.

\bibitem{Dahia:2015xxa}
F.~Dahia and A.~S. Lemos, \emph{{Constraints on extra dimensions from atomic spectroscopy}}, \href{https://doi.org/10.1103/PhysRevD.94.084033}{\emph{Phys. Rev. D} {\bfseries 94} (2016) 084033}, [\href{https://arxiv.org/abs/1509.06817}{{\ttfamily 1509.06817}}].

\bibitem{Dahia:2015bza}
F.~Dahia and A.~S. Lemos, \emph{{Is the proton radius puzzle evidence of extra dimensions?}}, \href{https://doi.org/10.1140/epjc/s10052-016-4266-7}{\emph{Eur. Phys. J. C} {\bfseries 76} (2016) 435}, [\href{https://arxiv.org/abs/1509.08735}{{\ttfamily 1509.08735}}].

\bibitem{landauQED}
V.~B. Berestetskii, E.~M. Lifshitz and L.~P. Pitaevskii, \emph{{Quantum Electrodynamics}}, vol.~4 of \emph{Course of Theoretical Physics}.
\newblock Pergamon Press, Oxford, 1982.

\bibitem{Biedenharn_Louck_1984}
L.~C. Biedenharn and J.~D. Louck, \emph{Angular Momentum in Quantum Physics: Theory and Application}.
\newblock Encyclopedia of Mathematics and its Applications. Cambridge University Press, 1984.

\bibitem{rose1961relativistic}
M.~E. Rose, \emph{Relativistic Electron Theory}.
\newblock John Wiley \& Sons, New York, 1961.

\bibitem{Szmytkowski:2007mzc}
R.~Szmytkowski, \emph{{Recurrence and differential relations for spherical spinors}}, \href{https://doi.org/10.1007/s10910-006-9110-0}{\emph{J. Math. Chem.} {\bfseries 42} (2007) 397--413}, [\href{https://arxiv.org/abs/1011.3433}{{\ttfamily 1011.3433}}].

\bibitem{sympy}
A.~Meurer, C.~P. Smith, M.~Paprocki, O.~\v{C}ert\'{i}k, S.~B. Kirpichev, M.~Rocklin et~al., \emph{Sympy: symbolic computing in python}, \href{https://doi.org/10.7717/peerj-cs.103}{\emph{PeerJ Computer Science} {\bfseries 3} (2017) e103}.

\bibitem{mpmath}
F.~Johansson, \emph{mpmath: a {P}ython library for arbitrary-precision floating-point arithmetic (version 1.3.0)},  2023, \href{https://mpmath.org}{https://mpmath.org}.

\bibitem{Kozlov:2018mbp}
M.~G. Kozlov, M.~S. Safronova, J.~R. Crespo L\'opez-Urrutia and P.~O. Schmidt, \emph{{Highly charged ions: Optical clocks and applications in fundamental physics}}, \href{https://doi.org/10.1103/RevModPhys.90.045005}{\emph{Rev. Mod. Phys.} {\bfseries 90} (2018) 045005}, [\href{https://arxiv.org/abs/1803.06532}{{\ttfamily 1803.06532}}].

\bibitem{Delaunay:2022grr}
C.~Delaunay, J.-P. Karr, T.~Kitahara, J.~C.~J. Koelemeij, Y.~Soreq and J.~Zupan, \emph{{Self-Consistent Extraction of Spectroscopic Bounds on Light New Physics}}, \href{https://doi.org/10.1103/PhysRevLett.130.121801}{\emph{Phys. Rev. Lett.} {\bfseries 130} (2023) 121801}, [\href{https://arxiv.org/abs/2210.10056}{{\ttfamily 2210.10056}}].

\bibitem{NIST_ASD}
A.~Kramida, {Yu.~Ralchenko}, J.~Reader and {and NIST ASD Team}. {NIST Atomic Spectra Database (ver. 5.11), [Online]. Available: {\tt{https://physics.nist.gov/asd}} [2024, July 15]. National Institute of Standards and Technology, Gaithersburg, MD.}, 2023.

\bibitem{10.1063/1.1796671}
O.~Jitrik and C.~F. Bunge, \emph{{Transition Probabilities for Hydrogen-Like Atoms}}, \href{https://doi.org/10.1063/1.1796671}{\emph{Journal of Physical and Chemical Reference Data} {\bfseries 33} (01, 2005) 1059--1070}.

\bibitem{Rostagni:2023eic}
G.~Rostagni and J.~D. Shergold, \emph{{The dark Stodolsky effect: constraining effective dark matter operators with spin-dependent interactions}}, \href{https://doi.org/10.1088/1475-7516/2023/07/018}{\emph{JCAP} {\bfseries 07} (2023) 018}, [\href{https://arxiv.org/abs/2304.06750}{{\ttfamily 2304.06750}}].

\bibitem{Read:2014qva}
J.~I. Read, \emph{{The Local Dark Matter Density}}, \href{https://doi.org/10.1088/0954-3899/41/6/063101}{\emph{J. Phys. G} {\bfseries 41} (2014) 063101}, [\href{https://arxiv.org/abs/1404.1938}{{\ttfamily 1404.1938}}].

\end{thebibliography}\endgroup



\end{document}